\documentclass[onecolumn,notitlepage,superscriptaddress]{revtex4-1} 
\usepackage[colorlinks=true,urlcolor=blue,linkcolor=blue,citecolor=blue]{hyperref}
\usepackage{latexsym}
\usepackage{graphicx}
\usepackage{subfigure} 
\usepackage{placeins}
\usepackage{amssymb}
\usepackage{amsmath}
\usepackage{color}
\usepackage{amsfonts}
\usepackage{mathrsfs}

\usepackage{accents}
\makeatletter
\def\wideubar{\underaccent{{\cc@style\underline{\mskip10mu}}}}
\def\Wideubar{\underaccent{{\cc@style\underline{\mskip8mu}}}}
\makeatother
\makeatletter
\def\widebar{\accentset{{\cc@style\underline{\mskip10mu}}}}
\def\Widebar{\accentset{{\cc@style\underline{\mskip8mu}}}}
\makeatother

\begin{document}

\title{ Post-experiment coincidence detection techniques for direct detection of two-body correlations }

\author{Dezhong Cao}
\affiliation{ Department of Physics, Yantai University, Yantai 264005, People's Republic of China }

\author{Yuehua Su}
\email{suyh@ytu.edu.cn}
\affiliation{ Department of Physics, Yantai University, Yantai 264005, People's Republic of China }

\begin{abstract}

It is one challenge to develop experimental techniques for direct detection of the many-body correlations of strongly correlated electrons, which exhibit a variety of unsolved mysteries. In this article, we present a \textit{post-experiment} coincidence counting method and propose two \textit{post-experiment} coincidence detection techniques, \textit{post-experiment} coincidence angle-resolved photoemission spectroscopy (cARPES) and \textit{post-experiment} coincidence inelastic neutron scattering (cINS). By coincidence detection of two photoelectric processes or two neutron-scattering processes, the \textit{post-experiment} coincidence detection techniques can detect directly the two-body correlations of strongly correlated electrons in particle-particle channel or two-spin channel. The \textit{post-experiment} coincidence detection techniques can be implemented upon the \textit{pulse}-resolved angle-resolved photoemission spectroscopy (ARPES) or inelastic neutron scattering (INS) experimental apparatus with \textit{pulse} photon or neutron source. When implemented experimentally, they will be powerful techniques to study the highly esoteric high-temperature superconductivity and the highly coveted quantum spin liquids.  

\end{abstract} 

\maketitle

\section{Introduction} \label{sec1}

In the field of condensed matter physics, it is one challenge to develop experimental techniques to study the many-body physics of strongly correlated electrons which are beyond the traditional theories \citep{StewartNFLRMP2001, PALeeRMP2006, StewartFeSCRMP2011, ChenXHNAR2014, FradkinARCMP2010, MengNature2010, BalentsNature2010, ZhouYiRMP2017}. Recently, some coincidence detection techniques have been proposed for direct detection of the two-body correlations of strongly correlated electrons \citep{SuZhang2020, DevereauxPRB2023, SucINS2021}. The basic principle of the proposed coincidence detection techniques is to utilize the second-order perturbations of the interaction between the target matter and the external probe field to detect the two-body responses of the target matter. By coincidence detection of two photoelectric processes which stem from the second-order perturbations of the electron-photon interaction, the coincidence angle-resolved photoemission spectroscopy (cARPES) can detect directly the two-body correlations of the target electrons in particle-particle channel \citep{SuZhang2020, DevereauxPRB2023}. Therefore, the cARPES can be developed to study the unconventional superconductivity \citep{GrusdtNC2024,Grusdt2023Feshbach}. Similarly, by coincidence detection of two neutron-scattering processes which come from the second-order perturbations of the electron-neutron spin interaction, the coincidence inelastic neutron scattering (cINS) can detect directly the two-spin correlations of the target electrons \citep{SucINS2021}. Thus, the cINS can be developed to investigate the novel quantum spin liquids \citep{TrivediPRB2023,WangTrivedi202403,ZhuTrivedi202405}. 

\begin{figure}[ht]
\centering
\includegraphics[width=0.5\columnwidth]{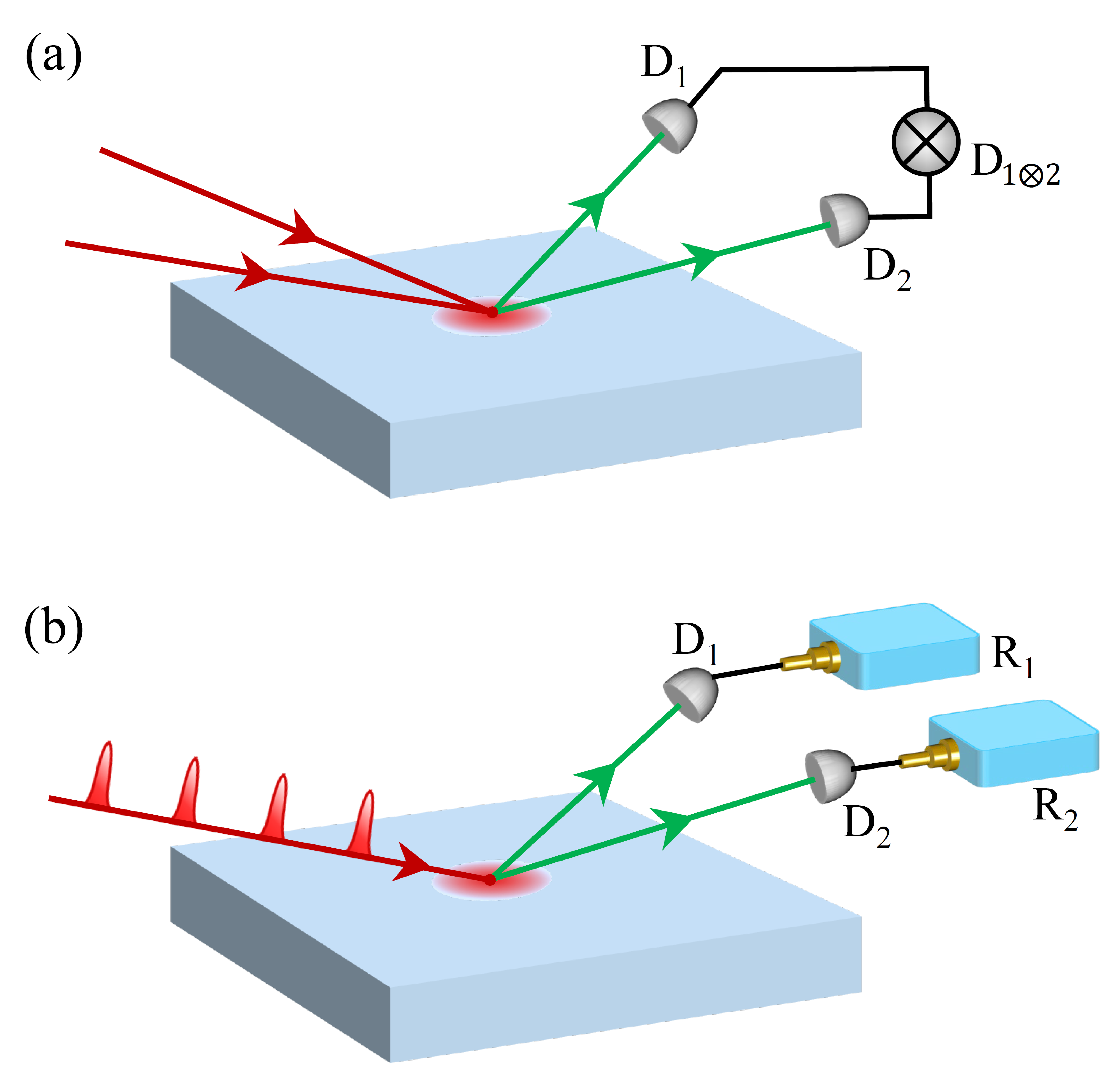} 
\caption{ Schematic illustration of the coincidence detection techniques. (a) The {\it instantaneous} coincidence detection technique with a coincidence detector $D_{1\otimes 2}$, (b) the {\it post-experiment} coincidence detection technique with a {\it pulse} source and two counting recorders $R_1$ and $R_2$. Here $D_1$ and $D_2$ are two single-photoelectron or single-neutron detectors. Fig. 1 (a) adapted with permission from Refs. \citep{SuZhang2020, SucINS2021}, copyrighted by the American Physical Society.}
\label{fig1}
\end{figure}

The original proposals for the coincidence detection techniques are schematically illustrated in Fig. \ref{fig1} (a). In the original proposal for the cARPES \citep{SuZhang2020}, two incident photons excite two photoelectrons which are detected by two single-photoelectron detectors $D_1$ and $D_2$, respectively. An additional coincidence detector $D_{1\otimes 2}$ records the coincidence counting of the emitted photoelectrons which arrive at these two detectors, thus recording the coincidence probability of two relevant photoelectric processes. The cINS is designed similarly to detect the coincidence probability of two neutron-scattering processes \citep{SucINS2021}. These originally proposed coincidence detection techniques can be named {\it instantaneous} coincidence detection techniques because the coincidence detector can record the coincidence probability {\it instantaneously} in experiment. It should be remarked that in our original proposals \citep{SuZhang2020, SucINS2021}, the coincidence detector $D_{1\otimes 2}$ makes once coincidence counting when the two detectors $D_1$ and $D_2$ each detect one photoelectron or one scattered neutron at {\it simultaneous} time. As the coincidence probability is defined for two photoelectric processes or two neutron-scattering processes which have finite occurrence time window, the coincidence counting made by the coincidence detector at exact {\it simultaneous} time is scientifically unnecessary and impossibly implemented in experiment. In order to perform coincidence detection of two photoelectric processes or two neutron-scattering processes with finite occurrence time window, the incident photons or neutrons can be designed to come from one {\it pulse} source. In this case, a time-window controller can be introduced in order for the coincidence detector to be able to perform coincidence counting of two photoelectric processes or two neutron-scattering processes caused by each incident photon or neutron {\it pulse}. 

In this article, we present a {\it post-experiment} coincidence counting method and propose two {\it post-experiment} coincidence detection techniques without a coincidence detector, {\it post-experiment} cARPES and {\it post-experiment} cINS. They can be implemented upon the {\it pulse}-resolved angle-resolved photoemission spectroscopy (ARPES) or inelastic neutron scattering (INS) experimental apparatus with {\it pulse} photon or neutron source. By developing an $S$-matrix perturbation theory, we show that the {\it post-experiment} coincidence detection techniques can obtain the coincidence probability of {\it pulse}-resolved two photoelectric processes or two neutron-scattering processes from a {\it post-experiment} coincidence counting method more easily and more efficiently than the {\it instantaneous} coincidence detection techniques. Since the coincidence probability involves the two-body correlations of the target electrons, the {\it post-experiment} coincidence detection techniques will be powerful techniques to study the various unconventional physics of strongly correlated electrons. 

\section{Post-experiment coincidence detection techniques} \label{sec2}

\subsection{Post-experiment coincidence counting method} \label{sec2.1}

The proposed experimental apparatus of the {\it post-experiment} coincidence detection techniques is schematically shown in Fig. \ref{fig1} (b). Let us first consider the {\it post-experiment} cARPES. Suppose the incident photons come from a {\it pulse} source. At times $t_n = t_0 + n \, \Delta t_d $ with $n=0, 1, 2, \cdots, N$, the photon source emits photon pulses sequentially, where $\Delta t_d$ is the time window between sequential two pulses. Each photon pulse is in a multiphoton state, which can cause many photoelectric processes. Suppose at time $t_{n-1}$, the photon source emits one photon pulse. At the same time, two counting recorders $R_1$ and $R_2$ begin to record the emitted photoelectrons which arrive at two single-photoelectron detectors $D_1$ and $D_2$, respectively. The $n$-th counting time is over before the beginning of the sequential next photon pulse in order that the two photoelectric processes caused by the time-$t_{n-1}$ pulse can be distinctly resolved. Define two variables, $I_{d_1}^{(1)}$ and  $I_{d_2}^{(1)}$, for the recorded counting data in the two respective recorders $R_1$ and $R_2$. Thus, we have two sequential recorded counting data, $\{ a_1(n), n=1, 2, \cdots, N\}$ for $I_{d_1}^{(1)}$ and $\{ a_2(n), n=1, 2, \cdots, N\}$ for $I_{d_2}^{(1)}$. Here $a_{1}(n), \, a_{2}(n) = 0$ or $1$ \citep{remark1}. This can be schematically shown in Table \ref{tab1}. With these pulse-resolved recorded data, we will introduce the following coincidence counting method. The coincidence counting of the $n$-th pair, $a_1(n)$ and $a_2(n)$, is defined by $I_d^{(2)} (n) = a_1(n)\times a_2(n)$, which defines the coincidence counting of the photoelectrons arrived at two single-photoelectron detectors $D_1$ and $D_2$ within the $n$-th time window $t\in [t_{n-1}, t_n)$ and thus describes the coincidence probability of the two photoelectric processes within this time window. It should be noted that when $a_1(n) = 1$ and $a_2(n) = 1$, $I_d^{(2)} (n) = 1$ which plays the same role of the coincidence detector of the {\it instantaneous} cARPES for the coincidence counting. 
The statistical average of the coincidence counting is defined by $\langle I_d^{(2)} \rangle = \frac{1}{N} \sum_{n=1}^{N} a_1(n) \times a_2(n)$. It involves the two-body correlations of the target electrons in particle-particle channel. Define another two statistical averages,  $\langle I_{d_1}^{(1)} \rangle =  \frac{1}{N} \sum_{n=1}^{N} a_1(n)$ and $\langle I_{d_2}^{(1)} \rangle = \frac{1}{N} \sum_{n=1}^{N} a_2(n)$. The intrinsic two-body correlations can be obtained by $ I_d^{(2,c)} = \langle I_d^{(2)} \rangle - \langle I_{d_1}^{(1)} \rangle \times \langle I_{d_2}^{(1)} \rangle$. This is a {\it post-experiment} cARPES coincidence detection technique. All of the above discussions can be similarly made for the cINS, thus we can also have a {\it post-experiment} cINS coincidence detection technique. 

One more remark is given on three time scales, $t_c$ the characteristic time scale of the physics we are interested in, $\Delta t_p$ the time width of the pulse, and $\Delta t_d$ the time window between sequential two pulses. In order to study the dynamics of the physics we are interested in, we should choose $\Delta t_p \leq t_c $ and $\Delta t_d \gg t_c$, which also ensures $\Delta t_d \gg \Delta t_p$ so that the two photoelectric processes or the two neutron-scattering processes from each incident pulse for one coincidence detection can be distinctly resolved \citep{remark2}. 


\begin{table}[h]
\centering
\begin{tabular}{|c|c|c|c|} 
\hline
Time window & $I_{d_1}^{(1)}$ & $I_{d_2}^{(1)}$ & $I_{d}^{(2)}$ \\ \hline 
$1$ & $a_1(1)$ & $a_2(1)$ & $a_1(1)\times a_2(1)$ \\ \hline
$2$ & $a_1(2)$ & $a_2(2)$ & $a_1(2)\times a_2(2)$ \\ \hline
\vdots & \vdots & \vdots & \vdots \\ \hline
$N$ & $a_1(N)$ & $a_2(N)$ & $a_1(N)\times a_2(N)$ \\ \hline
Average & $\langle I_{d_1}^{(1)} \rangle $ &  $\langle I_{d_2}^{(1)} \rangle $ &  $\langle I_{d}^{(2)} \rangle $ \\
\hline
\end{tabular}
\caption{ Post-experiment coincidence counting method. $I_{d_1}^{(1)}$ and $I_{d_2}^{(1)}$ are two variables defined for the two respective recorders $R_1$ and $R_2$, which record the counting data $a_1(n)$ and $a_2(n)$ within the $n$-th time window. $I_{d}^{(2)}$ defines the $n$-th coincidence counting. Three statistical averages are defined in the main text. }\label{tab1}
\end{table}

The above {\it post-experiment} coincidence counting method is based upon the following coincidence probability expression: 
\begin{eqnarray}
\widebar{\Gamma}^{(2)} = \Gamma^{(2)} \cdot I^{(2)}_\chi	\cdot I_d^{(2)} , \label{eqn1} 
\end{eqnarray} 
where $\Gamma^{(2)}$ is the coincidence probability obtained previously for the cARPES \citep{SuZhang2020} or the cINS \citep{SucINS2021} which can be regarded as a two-body correlation relevant target-electron form factor, $I^{(2)}_\chi$ defines an incident-particle-state factor, and 
\begin{equation}
I_d^{(2)} = I_{d_1}^{(1)} \times I_{d_2}^{(1)} \label{eqn2}
\end{equation} 
defines an emitted- or scattered-particle-state factor. It is $I_d^{(2)} = I_{d_1}^{(1)} \times I_{d_2}^{(1)}$ that makes the {\it post-experiment} coincidence counting method scientifically reasonable. In the below, we will show that the {\it post-experiment} cARPES and cINS coincidence detection techniques follow Eq. (\ref{eqn1}). 

\subsection{Post-experiment $\text{c}$ARPES} \label{sec2.2} 

Let us first consider the {\it post-experiment} cARPES following the reference \citep{SuZhang2020}. Suppose the electron-photon interaction \citep{SuZhang2020,Bruus2002} relevant to the photoelectric processes is defined by $V_{A} = \sum_{\mathbf{k}\sigma\mathbf{q}\lambda} g_A(\mathbf{k};\mathbf{q},\lambda) d^{\dag}_{\mathbf{k}+\mathbf{q}\sigma} c_{\mathbf{k}\sigma} a_{\mathbf{q}\lambda}$,
where $d^{\dag}_{\mathbf{k}\sigma}$ is the creation operator for the photoelectrons with momentum $\mathbf{k}$ and spin $\sigma$, $c_{\mathbf{k}\sigma}$ is the annihilation operator for the electrons in the target matter, $a_{\mathbf{q}\lambda}$ is the annihilation operator for the photons with momentum $\mathbf{q}$ and polarization $\lambda$. Introduce the electron-photon interaction relevant $S$-matrix $S_A=T_t \exp [-\frac{i}{\hbar} \int_{-\infty}^{+\infty} dt \, V_{A,I}(t) \cdot F(t)]$, where $V_{A,I}(t)= e^{i H_{A,0} t/\hbar} V_A e^{-i H_{A,0} t/\hbar}$. Here $T_t$ is a time-ordering operator, $H_{A,0}$ includes the Hamiltonians of the target electrons, the incident photons and the emitted photoelectrons, and $F(t)=\theta(t+\Delta t_d/2) - \theta(t-\Delta t_d/2)$ defines one time window where $\theta$ is the step function.

Suppose the incident photons from the pulse source have momentum $\mathbf{q}$ and polarization $\lambda$ with a distribution function $P_{A}(\mathbf{q},\lambda)$ and the emitted photoelectrons are focused with fixed momentum $\mathbf{k}$ and spin $\sigma$. The photoemission probability of one single-photoelectric process is defined by 
\begin{eqnarray}
\widebar{\Gamma}_{A,IF}^{(1)} = \big| \langle \Phi^{(1)}_{A,F} | S_A^{(1)}  | \Phi^{(1)}_{A,I} \rangle \big|^2 , \label{eqn3}
\end{eqnarray}  
where $S_A^{(1)}$ is the first-order expansion of the $S_A$ matrix, $|\Phi^{(1)}_{A,I} \rangle = | \Psi_{\alpha} \rangle \otimes | \chi_i(\mathbf{q} \lambda)\rangle \otimes | 0^{(d)} \rangle $ is the initial state and  $|\Phi^{(1)}_{A,F} \rangle = | \Psi_{\beta} \rangle \otimes | \chi_f(\mathbf{q} \lambda)\rangle \otimes | n^{(d)}_{\mathbf{k}\sigma} \rangle $ is the final state. Here $| \Psi_{\alpha} \rangle$ and $| \Psi_{\beta} \rangle$ are the target-electron eigenstates with the respective eigenenergies $E_{\alpha}$ and $E_{\beta}$, $|\chi_i(\mathbf{q} \lambda)\rangle$ and $|\chi_f(\mathbf{q} \lambda)\rangle$ are the photon initial and final states, and $n_{\mathbf{k}\sigma}^{(d)} = 0$ or $1$ is the photoelectron number defined for the photoelectron states. It should be remarked that $\widebar{\Gamma}_{A,IF}^{(1)}$ defines the photoemission probability of {\it once} single-photoelectric process in realistic ARPES measurement. It can be shown that
\begin{equation}
\widebar{\Gamma}_{A,IF}^{(1)} = \Gamma_{A,\alpha\beta}^{(1)} \cdot I_{A,\chi}^{(1)} \cdot I_{A,d}^{(1)} ,  \label{eqn4}
\end{equation}
where $\Gamma_{A,\alpha\beta}^{(1)}$ is a target-electron form factor, $I_{A,\chi}^{(1)}$ is a photon-state factor and $I_{A,d}^{(1)}$ is a photoelectron-state factor, the latter two of which are defined by
\begin{eqnarray}
I_{A,\chi}^{(1)} &=& \big| \langle \chi_f(\mathbf{q}\lambda) | a_{\mathbf{q}\lambda} | \chi_i(\mathbf{q}\lambda) \rangle \big|^2 , \notag \\
I_{A,d}^{(1)} &=& \big| \langle n^{(d)}_{\mathbf{k}\sigma} | d^\dag_{\mathbf{k}\sigma} | 0^{(d)} \rangle \big|^2 . \label{eqn5} 
\end{eqnarray}
Note that $I_{A,d}^{(1)} = 0 \, (1)$ when $n_{\mathbf{k}\sigma}^{(d)} = 0 \, (1)$. Therefore, the photoelectron-state factor plays a role to record the number of the photoelectrons arrived at the single-photoelectron detector.

The statistical average of the photoemission probability is shown to follow
\begin{equation}
\widebar{\Gamma}_A^{(1)} = \sum_{I F} P_{A}(\mathbf{q},\lambda) \cdot \Gamma_A^{(1)} \cdot I_{A,\chi}^{(1)} \cdot I_{A,d}^{(1)} ,  \label{eqn6}
\end{equation}
where $\sum_{I F} = \sum_{\mathbf{q}\lambda \chi_i \chi_f n^{(d)}}$, and $\Gamma_A^{(1)} = \frac{1}{Z}\sum_{\alpha\beta} e^{-\beta E_\alpha} \Gamma_{A,\alpha\beta}^{(1)}$ defines the photoemission probability of the ARPES obtained previously \citep{SuZhang2020}, 
\begin{equation}
\Gamma_A^{(1)} = \frac{|g_A|^{2}\Delta t_d}{\hbar} A(\mathbf{k}-\mathbf{q},\sigma; E_A^{(1)}) \cdot n_F(E_A^{(1)}) .  \label{eqn7}
\end{equation}
Here $A(\mathbf{k},\sigma; E) = -2 \, \text{Im}\, G_\sigma(\mathbf{k}, i\omega_n \rightarrow E+i\delta^{+})$ is the spectral function of the imaginary-time Green's function $G_\sigma(\mathbf{k},\tau) = - \langle T_\tau c_{\mathbf{k}\sigma}(\tau) c_{\mathbf{k}\sigma}^\dag (0)  \rangle$, $n_F(E)$ is the Fermi-Dirac distribution function, $g_A = g_A(\mathbf{k}-\mathbf{q};\mathbf{q},\lambda)$, and $E_A^{(1)}$ is the transferred energy in the photoelectric process. $E_A^{(1)}$ is defined by $E_A^{(1)} = \varepsilon_{\mathbf{k}}^{(d)}+\Phi-\hbar \omega_{\mathbf{q}}$, where $\varepsilon_{\mathbf{k}}^{(d)}$ is the photoelectron energy, $\Phi$ is the work function, and $\hbar\omega_{\mathbf{q}}$ is the photon energy. 

It should be remarked that the photoelectron-state factor $I_{A,d}^{(1)}$ makes us to obtain the {\it absolute} counting of the photoemission probability in realistic ARPES measurement, with zero counting when $n^{(d)}_{\mathbf{k}\sigma}=0$ and $I_{A,d}^{(1)}=0$ and finite counting when $n^{(d)}_{\mathbf{k}\sigma}=1$ and $I_{A,d}^{(1)}=1$. This is different from the conventional ARPES measurement, where only the signals with $n^{(d)}_{\mathbf{k}\sigma}=1$ and $I_{A,d}^{(1)}=1$ are recorded and only the {\it relative} photoemission probability can be obtained. Moreover, as shown in the below, it is the photoelectron-state factors that make the {\it post-experiment} cARPES scientifically correct and experimentally realizable.  

Let us now consider the coincidence detection of two photoelectric processes caused by one incident photon pulse for the {\it post-experiment} cARPES, where the incident photons have same momentum and polarization $(\mathbf{q},\lambda)$ and the photoelectrons arrived at two single-photoelectron detectors have fixed momenta and spins $(\mathbf{k}_1  \sigma_1)$ and $(\mathbf{k}_2 \sigma_2)$, respectively. The case where the incident photons have different momenta and polarizations can be discussed with a similar procedure given below. The coincidence probability of the two photoelectric processes is defined by 
\begin{eqnarray}
\widebar{\Gamma}_{A,IF}^{(2)} = \big| \langle \Phi^{(2)}_{A,F} | S_A^{(2)}  | \Phi^{(2)}_{A,I} \rangle \big|^2 , \label{eqn8}
\end{eqnarray}  
where $S_A^{(2)}$ is the second-order expansion of the $S_A$ matrix \citep{SuZhang2020}, $|\Phi^{(2)}_{A,I} \rangle$ and $|\Phi^{(2)}_{A,F} \rangle$ are the corresponding initial and final states which are defined by $|\Phi^{(2)}_{A,I} \rangle = | \Psi_{\alpha} \rangle \otimes | \chi_i(\mathbf{q} \lambda)\rangle \otimes | 0^{(d)} \rangle $ and $|\Phi^{(2)}_{A,F} \rangle = | \Psi_{\beta} \rangle \otimes | \chi_f(\mathbf{q} \lambda)\rangle \otimes | n^{(d)}_{\mathbf{k}_1\sigma_1} n^{(d)}_{\mathbf{k}_2\sigma_2} \rangle $. $\widebar{\Gamma}_{A,IF}^{(2)}$ can be shown to follow
\begin{equation}
\widebar{\Gamma}_{A,IF}^{(2)} = \Gamma_{A,\alpha\beta}^{(2)} \cdot I_{A,\chi}^{(2)} \cdot I_{A,d}^{(2)} ,  \label{eqn9}
\end{equation}
where the target-electron form factor $\Gamma_{A,\alpha\beta}^{(2)}$ follows
\begin{equation}
\Gamma_{A,\alpha\beta}^{(2)} = \frac{|g_{A,1}g_{A,2}|^2}{\hbar^4}  \big| \Phi_{A,\alpha\beta}^{(2)} (\mathbf{k}_{A,1} \sigma_1,\mathbf{k}_{A,2} \sigma_2;\Omega_A,\omega_A) \big|^2   \label{eqn10}
\end{equation}
with $\mathbf{k}_{A,1} = \mathbf{k}_1-\mathbf{q}$, $\mathbf{k}_{A,2} = \mathbf{k}_2-\mathbf{q}$, $g_{A,1}= g_A(\mathbf{k}_{A,1};\mathbf{q},\lambda)$ and $g_{A,2} = g_A(\mathbf{k}_{A,2};\mathbf{q},\lambda)$. Here in order to describe the coincidence probability $\Gamma_{A,\alpha\beta}^{(2)}$, we have introduced  a two-body Bethe-Salpeter wave function in particle-particle channel \citep{SalpeterBethe1951, GellmanLowBS1951},
\begin{equation}
\Phi^{(2)}_{A,\alpha\beta}(\mathbf{k}_1 \sigma_1 t_1; \mathbf{k}_2 \sigma_2 t_2) = \langle \Psi_{\beta} | T_t c_{\mathbf{k}_2\sigma_2}(t_2) c_{\mathbf{k}_1\sigma_1} (t_1) | \Psi_{\alpha }\rangle . \label{eqn11}
\end{equation}
Defining $t_c = (t_1+t_2)/2$ and $t_r = t_2 - t_1$, we can introduce another expression of the two-body Bethe-Salpeter wave function, $\Phi^{(2)}_{A,\alpha\beta}(\mathbf{k}_1 \sigma_1, \mathbf{k}_2 \sigma_2; t_c, t_r)=\Phi^{(2)}_{A,\alpha\beta}(\mathbf{k}_1 \sigma_1 t_1; \mathbf{k}_2 \sigma_2 t_2)$. $\Phi_{A,\alpha\beta}^{(2)} (\mathbf{k}_1\sigma_1,\mathbf{k}_2\sigma_2;\Omega,\omega)$ is the Fourier transformation of $\Phi^{(2)}_{A,\alpha\beta}(\mathbf{k}_1 \sigma_1, \mathbf{k}_2 \sigma_2; t_c, t_r)$ and defined as 
\begin{eqnarray}
\Phi^{(2)}_{A,\alpha\beta}(\mathbf{k}_1 \sigma_1, \mathbf{k}_2\sigma_2; \Omega, \omega) =  \iint_{-\infty}^{+\infty} d t_c d t_r \Phi^{(2)}_{A,\alpha\beta}(\mathbf{k}_1 \sigma_1, \mathbf{k}_2\sigma_2; t_c, t_r)  e^{i \Omega t_c + i \omega t_r} .  \qquad  \label{eqn12}
\end{eqnarray}
In Eq. (\ref{eqn10}), the center-of-mass frequency $\Omega_A$ and the relative frequency $\omega_A$ are defined by $\Omega_A =  (E_{A,1}+E_{A,2})/\hbar, \ \omega_A =  (E_{A,2}-E_{A,1})/2\hbar$, where the two transferred energies in the two photoelectric processes are defined by $E_{A,1} = \varepsilon_{\mathbf{k}_1}^{(d)} + \Phi - \hbar \omega_{\mathbf{q}}$ and $E_{A,2} = \varepsilon_{\mathbf{k}_2}^{(d)} + \Phi - \hbar \omega_{\mathbf{q}}$. In Eq. (\ref{eqn9}), the photon-state factor $I_{A,\chi}^{(2)}$ is defined by 
\begin{equation}
I_{A,\chi}^{(2)} = \big| \langle \chi_f(\mathbf{q}\lambda) | a_{\mathbf{q}\lambda}^2 | \chi_i(\mathbf{q}\lambda) \rangle \big|^2 , \label{eqn13}
\end{equation}
and the photoelectron-state factor $I_{A,d}^{(2)}$ is defined as
\begin{equation}
I_{A,d}^{(2)} = I_{A,d_1}^{(1)} \times  I_{A,d_2}^{(1)}, \label{eqn14}
\end{equation}
where 
\begin{eqnarray}
I_{A,d_1}^{(1)} &=& \big| \langle n^{(d)}_{\mathbf{k}_1\sigma_1} | d^\dag_{\mathbf{k}_1\sigma_1} | 0^{(d)} \rangle \big|^2, \notag \\
I_{A,d_2}^{(1)} &=& \big| \langle n^{(d)}_{\mathbf{k}_2\sigma_2} | d^\dag_{\mathbf{k}_2\sigma_2} | 0^{(d)} \rangle \big|^2 . \label{eqn15} 
\end{eqnarray}
Since $I_{A,d_1}^{(1)} = 0 \, (1)$ when $n_{\mathbf{k}_1\sigma_1}^{(d)} = 0 \, (1)$ and $I_{A,d_2}^{(1)} = 0 \, (1)$ when $n_{\mathbf{k}_2\sigma_2}^{(d)} = 0 \, (1)$, $I_{A,d}^{(2)}$ records the coincidence counting of the {\it pulse}-resolved photoelectrons arrived at two single-photoelectron detectors $D_1$ and $D_2$. 

The statistical average of the coincidence probability of {\it pulse}-resolved two photoelectric processes from every one of the sequential photon pulses is given by 
\begin{equation}
\widebar{\Gamma}_A^{(2)} = \frac{1}{Z}\sum_{I F} e^{-\beta E_\alpha} P_{A}(\mathbf{q},\lambda) \cdot \Gamma_{A,\alpha\beta}^{(2)} \cdot I_{A,\chi}^{(2)} \cdot I_{A,d}^{(2)} ,  \label{eqn16}
\end{equation}
where $\sum_{I F} = \sum_{\alpha\beta} \sum_{\mathbf{q}\lambda \chi_i \chi_f } \sum_{n_1^{(d)} n_2^{(d)}}$. It should be remarked that $\widebar{\Gamma}_{A,IF} ^{(2)}$ has a same structure to $\widebar{\Gamma}^{(2)}$ in Eq. (\ref{eqn1}) and $I_{A,d}^{(2)} = I_{A,d_1}^{(1)} \times  I_{A,d_2}^{(1)}$ follows in both $\widebar{\Gamma}_{A,IF} ^{(2)}$ and $\widebar{\Gamma}_A^{(2)}$. This shows that the coincidence probability of {\it pulse}-resolved two photoelectric processes can be obtained by $I_{A,d}^{(2)}$ which records the coincidence counting of the {\it pulse}-resolved photoelectrons arrived at two single-photoelectron detectors renormalized by the target-electron form factor and the photon-state factor. Therefore, a {\it post-experiment} cARPES can be designed following the {\it post-experiment}  coincidence counting method we have presented above. It is noted that when $I_{A,d_1}^{(1)} = 1$ and $I_{A,d_2}^{(1)} = 1$, $\widebar{\Gamma}_A^{(2)}$ can recover the previous results we have obtained for the {\it instantaneous} cARPES \citep{SuZhang2020}.  

It should be remarked that the coincidence probability of the {\it post-experiment} cARPES measurement is an {\it absolute} coincidence probability of two photoelectric processes. It is different from the {\it relative} one of the {\it instantaneous} cARPES measurement \citep{SuZhang2020} where only the coincidence detection signals with $I_{A,d_1}^{(1)} = 1$ and $I_{A,d_2}^{(1)} = 1$ are recorded. This difference comes from the introduction of a photoelectron-state factor $I_{A,d}^{(2)}=I_{A,d_1}^{(1)}\times I_{A,d_2}^{(1)}$ in the {\it post-experiment} cARPES. Moreover, the {\it post-experiment} cARPES can make the {\it post-experiment} coincidence counting more easily and more efficiently for the coincidence probability of {\it any} two photoelectric processes. Suppose we have obtained the photoemission counting data for many {\it pulse}-resolved photoelectric processes, which are recorded in the recorders $R_1, R_2, \cdots, R_M$ with different focused photoelectron momenta and spins. We can obtain the coincidence probability of {\it any} two photoelectric processes relevant to {\it any} two recorders $R_i$ and $R_j$ by using the coincidence counting $I_{d,ij}^{(2)}=I_{d,i}^{(1)} \times I_{d,j}^{(1)}$. Therefore, the {\it post-experiment} cARPES will be a highly efficient technique to obtain the coincidence probability of {\it pulse}-resolved two photoelectric processes, and thus will be powerful technique for coincidence detection of the two-body correlations of the target electrons.

\subsection{Post-experiment $\text{c}$INS} \label{sec2.3}

All of the above discussions can be extended into the case of the {\it post-experiment} cINS. Suppose the electron-neutron spin interaction \citep{SucINS2021,Lovesey1984,Squires1996,FelixPrice2013} is given by $V_B = \sum_{\mathbf{q}_i \mathbf{q}_f \sigma_i \sigma_f} g_B(\mathbf{q}) f^{\dag}_{\mathbf{q}_f \sigma_f} \boldsymbol{\tau}_{\sigma_f \sigma_i} f_{\mathbf{q}_i \sigma_i} \cdot \mathbf{S}_{\perp}(\mathbf{q})$ with $\mathbf{q} = \mathbf{q}_f - \mathbf{q}_i$. Here $f^{\dag}_{\mathbf{q} \sigma}$ and $f_{\mathbf{q} \sigma}$ are the neutron creation and annihilation operators with momentum $\mathbf{q}$ and spin $\sigma$, $\boldsymbol{\tau}$ is the Pauli matrix, and $\mathbf{S}_{\perp}(\mathbf{q})$ is a target-electron spin relevant operator. $\mathbf{S}_{\perp}(\mathbf{q})$ is defined as $\mathbf{S}_{\perp}(\mathbf{q}) =  \mathbf{S}(\mathbf{q}) \cdot (1 - \widehat{\mathbf{q}}\widehat{\mathbf{q}})$, where $\mathbf{S}(\mathbf{q}) = \sum_{l} \mathbf{S}_l e^{-i\mathbf{q}\cdot \mathbf{R}_l}$ with $\mathbf{S}_l$ being the target-electron spin operator at position $\mathbf{R}_l$ and $\widehat{\mathbf{q}} = \mathbf{q}/|\mathbf{q}|$. The electron-neutron scattering $S$-matrix is defined by $S_B=T_t \exp [-\frac{i}{\hbar} \int_{-\infty}^{+\infty} dt \, V_{B,I}(t) \cdot F(t)]$, where $V_{B,I}(t)= e^{i H_{B,0} t/\hbar} V_B e^{-i H_{B,0} t/\hbar}$. Here $H_{B,0}$ includes the Hamiltonians of the target-electron spin system and the neutrons. 

Consider one single neutron-scattering process of the INS measurement with the initial state $|\Phi^{(1)}_{B,I} \rangle = | \Psi_{\alpha} \rangle \otimes | n_{\mathbf{q}_i\sigma_i} \rangle $ and the final state $|\Phi^{(1)}_{B,F} \rangle = | \Psi_{\beta} \rangle \otimes | n_{\mathbf{q}_f\sigma_f} \rangle $. Here $n_{\mathbf{q}\sigma}=0$ or $1$ is the neutron number defined for the neutron states. The scattering probability of this single neutron-scattering process can be defined by $\widebar{\Gamma}_{B,IF}^{(1)} = \big| \langle \Phi^{(1)}_{B,F} | S_B^{(1)}  | \Phi^{(1)}_{B,I} \rangle \big|^2$, where $S_B^{(1)}$ is the first-order expansion of the $S_B$ matrix. Following the above procedure for the ARPES and the previous derivation for the INS \citep{SucINS2021}, we can show that 
\begin{equation}
\widebar{\Gamma}_{B,IF}^{(1)} = \Gamma_{B,\alpha\beta}^{(1)} \cdot I_{B,\chi}^{(1)} \cdot I_{B,d}^{(1)} ,  \label{eqn17}
\end{equation}
where $\Gamma_{B,\alpha\beta}^{(1)}$ is a target-electron spin form factor, $I_{B,\chi}^{(1)} = \big| \langle 0 | f_{\mathbf{q}_i\sigma_i} | n_{\mathbf{q}_i\sigma_i} \rangle \big|^2  $ defines an incident-neutron-state factor and $I_{B,d}^{(1)} = \big| \langle n_{\mathbf{q}_f\sigma_f} | f^\dag_{\mathbf{q}_f\sigma_f} | 0 \rangle \big|^2 $ defines a scattered-neutron-state factor. It should be noted that $I_{B,\chi}^{(1)} = 0 \, (1)$ when $n_{\mathbf{q}_i\sigma_i} = 0 \, (1)$ and $I_{B,d}^{(1)} = 0 \, (1)$ when $n_{\mathbf{q}_f\sigma_f} = 0 \, (1)$. The scattered-neutron-state factor plays a role to record the number of the scattered neutrons arrived at the single-neutron detector. 
Suppose the incident neutrons from the neutron pulses follow a distribution $P_{B}^{(1)}(\mathbf{q}_i, \sigma_i) = P_{B}^{(1)}(\mathbf{q}_i) \cdot \widebar{P}_{B}^{(1)}(\sigma_i)$, where the neutron spins are in the mixed states defined by $\sum_{\sigma_i} \widebar{P}_{B}^{(1)} (\sigma_i) \vert \sigma_i \rangle \langle \sigma_i \vert = \frac{1}{2} \left( \vert \uparrow \rangle \langle \uparrow \vert + \vert \downarrow \rangle \langle \downarrow \vert \right)$, and suppose the scattered neutrons which arrive at the single-neutron detector have fixed momentum $\mathbf{q}_f$ but arbitrary spin $\sigma_f$. The statistical average of the single neutron-scattering probability for the INS measurement can be shown to follow
\begin{equation}
\widebar{\Gamma}_B^{(1)} = \sum_{I F} P_{B}^{(1)}(\mathbf{q}_i) \cdot \Gamma_B^{(1)} \cdot I_{B,\chi}^{(1)} \cdot I_{B,d}^{(1)} ,  \label{eqn18}
\end{equation}
where $\sum_{I F} = \sum_{\mathbf{q}_i n_i n_{f}}$, and $\Gamma_B^{(1)}$ follows   
\begin{equation}
\Gamma_B^{(1)} = \frac{|g(\mathbf{q})|^{2} \Delta t_d}{\hbar} \chi_B(\mathbf{q}, E_B^{(1)}) \cdot n_B(E_B^{(1)}) .  \label{eqn19}
\end{equation}
Here $\chi_B(\mathbf{q},E)=-2 \, \text{Im}\, D(\mathbf{q}, i\nu_n \rightarrow E+i\delta^{+})$ is the spectral function of the target-electron spin Green's function $D(\mathbf{q},\tau) = - \sum_{ij} \langle T_\tau \boldsymbol{S}_i(\mathbf{q},\tau) \boldsymbol{S}_j^{\dag}(\mathbf{q},0) \rangle (\delta_{ij} - \widehat{\mathbf{q}}_i \widehat{\mathbf{q}}_j)$, $n_B(E)$ is the Bose-Einstein distribution function. The transferred energy $E_B^{(1)}=\mathcal{E}(\mathbf{q}_{f}) - \mathcal{E}(\mathbf{q}_{i})$, where $\mathcal{E}(\mathbf{q}_{i})$ and $\mathcal{E}(\mathbf{q}_{f})$ are respectively the incident and the scattered neutron energies. In the derivations of Eqs. (\ref{eqn18}) and (\ref{eqn19}), we have used the identity $\frac{1}{2}\sum_{\sigma_i \sigma_f } \langle \sigma_i|\boldsymbol{\tau}^{l} | \sigma_f \rangle \langle \sigma_f|\boldsymbol{\tau}^{l^\prime} | \sigma_i \rangle =\delta_{l l^\prime} $ for the non-polarized neutrons. 

Let us now consider the coincidence probability of {\it pulse}-resolved two neutron-scattering processes for the {\it post-experiment} cINS measurement following the reference \citep{SucINS2021}. For one coincidence detection with the initial state $|\Phi^{(2)}_{B,I} \rangle = | \Psi_{\alpha} \rangle \otimes | n_{\mathbf{q}_{i_1}\sigma_{i_1}} n_{\mathbf{q}_{i_2}\sigma_{i_2}} \rangle $ and the final state $|\Phi^{(2)}_{B,F} \rangle = | \Psi_{\beta} \rangle \otimes | n_{\mathbf{q}_{f_1}\sigma_{f_1}} n_{\mathbf{q}_{f_2}\sigma_{f_2}} \rangle $, the coincidence probability of the two neutron-scattering processes caused by one incident neutron pulse is  defined by $\widebar{\Gamma}_{B,IF}^{(2)} = \big| \langle \Phi^{(2)}_{B,F} | S_B^{(2)}  | \Phi^{(2)}_{B,I} \rangle \big|^2$, where $S_B^{(2)}$ is the second-order expansion of the $S_B$ matrix. It can be shown that  
\begin{equation}
\widebar{\Gamma}_{B,IF}^{(2)} = \Gamma_{B,\alpha\beta}^{(2)} \cdot I_{B,\chi}^{(2)} \cdot I_{B,d}^{(2)} ,  \label{eqn20}
\end{equation}
where $\Gamma_{B,\alpha\beta}^{(2)}$ is a target-electron spin form factor, $I_{B,\chi}^{(2)} = \big| \langle 0 | f_{\mathbf{q}_{i_1}\sigma_{i_1}} | n_{\mathbf{q}_{i_1}\sigma_{i_1}} \rangle \big|^2 \cdot \big| \langle 0 | f_{\mathbf{q}_{i_2}\sigma_{i_2}} | n_{\mathbf{q}_{i_2}\sigma_{i_2}} \rangle \big|^2$ is an incident-neutron-state factor, and $I_{B,d}^{(2)}$ is a scattered-neutron-state factor defined by 
\begin{equation}
I_{B,d}^{(2)} = I_{B,d_1}^{(1)} \times I_{B,d_2}^{(1)} , \label{eqn21}
\end{equation} 
where $ I_{B,d_1}^{(1)} = \big| \langle n_{\mathbf{q}_{f_1}\sigma_{f_1}} | f^\dag_{\mathbf{q}_{f_1}\sigma_{f_1}} | 0 \rangle \big|^2$ and $ I_{B,d_2}^{(1)} = \big| \langle n_{\mathbf{q}_{f_2}\sigma_{f_2}} | f^\dag_{\mathbf{q}_{f_2}\sigma_{f_2}} | 0 \rangle \big|^2 $. It is noted that $\widebar{\Gamma}_{B,IF}^{(2)}$ has a same structure to Eq. (\ref{eqn1}). Since $I_{B,d_1}^{(1)} = 0 \, (1)$ when $n_{\mathbf{q}_{f_1}\sigma_{f_1}} = 0 \, (1)$ and $I_{B,d_2}^{(1)} = 0 \, (1)$ when $n_{\mathbf{q}_{f_2}\sigma_{f_2}} = 0 \, (1)$, $I_{B,d}^{(2)}$ records the coincidence counting of the scattered neutrons arrived at two single-neutron detectors. 

Suppose the incident two neutrons from the sequential neutron pulses have momentum and spin distribution functions $P^{(2)}_{B}(\mathbf{q}_{i_1}, \mathbf{q}_{i_2}) = P^{(1)}_{B}(\mathbf{q}_{i_1}) \cdot P^{(1)}_{B}(\mathbf{q}_{i_2})$ and $\widebar{P}^{(2)}_{B}(\sigma_{i_1}, \sigma_{i_2}) = \widebar{P}^{(1)}_{B}(\sigma_{i_1}) \cdot \widebar{P}^{(1)}_{B}(\sigma_{i_2})$. Here $\widebar{P}_{B}^{(1)}(\sigma_{i})$ is defined as in the above INS case with the same neutron-spin mixed states. Suppose the two scattered neutrons are focused with fixed momenta $(\mathbf{q}_{f_1}, \mathbf{q}_{f_2} )$ but arbitrary spins $(\sigma_{f_1}, \sigma_{f_2})$. The statistical average of the  coincidence probability of {\it pulse-resolved} two neutron-scattering processes from every one of the sequential neutron pulses follows
\begin{equation}
\widebar{\Gamma}_B^{(2)} = \sum_{I F} P^{(2)}_{B}(\mathbf{q}_{i_1},\mathbf{q}_{i_2}) \cdot \Gamma_{B}^{(2)} \cdot I_{B,\chi}^{(2)} \cdot I_{B,d}^{(2)} ,  \label{eqn22}
\end{equation}
where $\sum_{I F} = \sum_{\mathbf{q}_{i_1} \mathbf{q}_{i_2}} \sum_{ n_{i_1} n_{i_2} n_{f_1} n_{f_2} }$, and $\Gamma_{B}^{(2)}$ is given by \citep{SucINS2021} 
\begin{equation}
\Gamma_{B}^{(2)} = \Gamma_{B,1}^{(2)} + \Gamma_{B,2}^{(2)} , \label{eqn23}
\end{equation}
with the two contributions defined as
\begin{eqnarray}
\Gamma_{B,1}^{(2)} &=& \frac{1}{Z}\sum_{\alpha\beta i j} e^{-\beta E_{\alpha}} C_1 \big\vert \phi^{(ij)}_{\alpha\beta}(\mathbf{q}_1 , \mathbf{q}_2; \Omega_B,\omega_B)  \big\vert^2 , \notag \\
\Gamma_{B,2}^{(2)} &=& \frac{1}{Z}\sum_{\alpha\beta i j} e^{-\beta E_{\alpha}} C_2 \big\vert \phi^{(ij)}_{\alpha\beta}(\overline{\mathbf{q}}_1 , \overline{\mathbf{q}}_2; \overline{\Omega}_B, \overline{\omega}_B) \big\vert^2 .  \label{eqn24} 
\end{eqnarray}
Here we have introduced a two-spin Bethe-Salpeter wave function which describes the two-spin correlations of the target electrons, 
\begin{equation}
\phi^{(ij)}_{\alpha\beta}(\mathbf{q}_1 t_1, \mathbf{q}_2 t_2) = \langle \Psi_\beta \vert T_t S^{(j)}_\perp (\mathbf{q}_2, t_2) S^{(i)}_\perp (\mathbf{q}_1, t_1) \vert \Psi_\alpha \rangle .  \label{eqn25}
\end{equation}
Similar to the definition of Eq. (\ref{eqn12}), $\phi^{(ij)}_{\alpha\beta}(\mathbf{q}_1 , \mathbf{q}_2; \Omega,\omega)$ is the Fourier transformation of $\phi^{(ij)}_{\alpha\beta}(\mathbf{q}_1, \mathbf{q}_2; t_c, t_r ) = \phi^{(ij)}_{\alpha\beta}(\mathbf{q}_1 t_1, \mathbf{q}_2 t_2)$ with $t_c = (t_1 + t_2)/2$ and $t_r = t_2 - t_1$. The two contributions, $\Gamma_{B,1}^{(2)}$ and $\Gamma_{B,2}^{(2)}$, come from two  different classes of microscopic neutron-scattering processes,  the former with the neutron-state changes as $\vert \mathbf{q}_{i_1}\sigma_{i_1}\rangle \rightarrow \vert \mathbf{q}_{f_1}\sigma_{f_1}\rangle$ and $\vert \mathbf{q}_{i_2}\sigma_{i_2}\rangle \rightarrow \vert \mathbf{q}_{f_2}\sigma_{f_2}\rangle$, and the latter with the neutron-state changes as $\vert \mathbf{q}_{i_1}\sigma_{i_1}\rangle \rightarrow \vert \mathbf{q}_{f_2}\sigma_{f_2}\rangle$ and $ \vert \mathbf{q}_{i_2}\sigma_{i_2}\rangle \rightarrow \vert \mathbf{q}_{f_1}\sigma_{f_1}\rangle$. In Eq. (\ref{eqn24}),  the transferred momenta are defined by $\mathbf{q}_1 = \mathbf{q}_{f_1}-\mathbf{q}_{i_1} , \mathbf{q}_2 = \mathbf{q}_{f_2}-\mathbf{q}_{i_2}, \overline{\mathbf{q}}_1 = \mathbf{q}_{f_1}-\mathbf{q}_{i_2} , \overline{\mathbf{q}}_2 = \mathbf{q}_{f_2}-\mathbf{q}_{i_1} $, and the transferred frequencies are defined by $ \Omega_B = (E_{B,1} + E_{B,2} )/\hbar, \omega_B = (E_{B,2} - E_{B,1} )/2\hbar, \overline{\Omega}_B = (\overline{E}_{B,1} + \overline{E}_{B,2} )/\hbar, \overline{\omega}_B = (\overline{E}_{B,2} - \overline{E}_{B,1} )/2\hbar$, where the transferred energies are defined as $E_{B,1} = \mathcal{E}(\mathbf{q}_{f_1}) - \mathcal{E}(\mathbf{q}_{i_1}) , E_{B,2} = \mathcal{E}(\mathbf{q}_{f_2}) - \mathcal{E}(\mathbf{q}_{i_2}) , \overline{E}_{B,1} = \mathcal{E}(\mathbf{q}_{f_1}) - \mathcal{E}(\mathbf{q}_{i_2}) ,  \overline{E}_{B,2} = \mathcal{E}(\mathbf{q}_{f_2}) - \mathcal{E}(\mathbf{q}_{i_1})$. The two constants $C_1$ and $C_2$ are given by $C_1 = \vert g_B(\mathbf{q}_1) g_B(\mathbf{q}_2) \vert^2/\hbar^4$ and $C_2 = \vert g_B(\overline{\mathbf{q}}_1) g_B(\overline{\mathbf{q}}_2) \vert^2/\hbar^4 $. 
It is noted that the coincidence probabilities $\widebar{\Gamma}_{B,IF}^{(2)}$ and $\widebar{\Gamma}_{B}^{(2)}$ both follow Eq. (\ref{eqn1}) with $I_{B,d}^{(2)} = I_{B,d_1}^{(1)} \times I_{B,d_2}^{(1)}$. Therefore, the coincidence probability of {\it pulse-resolved} two neutron-scattering processes can be obtained by $I_{B,d}^{(2)}$ with the renormalization of the target-electron spin form factor and the incident-neutron-state factor. It is clear that the cINS with a pulse neutron source can be designed into a {\it post-experiment} coincidence detection technique.


\section{Conclusion} \label{sec3}

In conclusion, we have shown that the {\it post-experiment} cARPES and cINS coincidence detection techniques follow Eq. (\ref{eqn1}). Therefore, the coincidence probability of {\it pulse}-resolved two photoelectric processes or two neutron-scattering processes can be obtained by these {\it post-experiment} coincidence detection techniques with the proposed {\it post-experiment} coincidence counting method. With a {\it pulse} photon or neutron source, the {\it post-experiment} coincidence detection techniques can be implemented upon the {\it pulse}-resolved ARPES or INS experimental apparatus. Since the coincidence probability of two photoelectric processes or two neutron-scattering processes involves the two-body correlations of the target electrons, the {\it post-experiment} coincidence detection techniques will be powerful techniques for investigating the various unsolved coveted mysteries of strongly correlated electrons. 

\section*{Acknowledgements}

We thank Prof. Yuan Li and Prof. Shan Qiao for invaluable discussions. This work was supported by the National Natural Science Foundation of China (Grants No. 11774299 and No. 11874318) and the Natural Science Foundation of Shandong Province (Grant No. ZR2023MA015).

\appendix

\section*{$S$-matrix perturbation theory for post-experiment coincidence detection techniques}

\subsection{A general $S$-matrix perturbation detection theory} \label{seca1.1}

Suppose at time $t\leq t_i$, a system is in thermodynamic equilibrium which can be described by a density matrix
\begin{equation}
\rho_0 = \frac{1}{Z} e^{-\beta H_0} , \label{eqnM1.1}
\end{equation}
where $H_0$ is a time-independent Hamiltonian and the partition function $Z = \text{Tr}(e^{-\beta H_0})$. At time $t_i$, a detection interaction $V(t)$ is turned on and the Hamiltonian becomes into the following form as 
\begin{equation}
H(t) = H_0(t) + V(t) , \label{eqnM1.2}
\end{equation} 
where $H_0(t)$ may become time dependent after $t_i$. Let us introduce two time-evolution operators $U_0 (t,t_i) = T_t \exp [-\frac{i}{\hbar}\int_{t_i}^t d t_1 H_0(t_1)]$ and $U_H (t,t_i) = T_t \exp [-\frac{i}{\hbar}\int_{t_i}^t d t_1 H (t_1)]$. An $S$-matrix in the interaction picture can be defined as $S(t,t_i) = U_0^{\dag}(t,t_i) U_H(t,t_i)$, which can be shown to follow
\begin{equation}
S(t,t_i) = T_t \exp[-\frac{i}{\hbar}\int_{t_i}^t d t_1 V_I (t_1)] , \label{eqnM1.3}
\end{equation} 
where $V_I(t)$ is the representation of $V(t)$ in the interaction picture and defined by 
\begin{equation}
V_I(t) = U_0^{\dag}(t,t_i) V(t) U_0(t,t_i) . \label{eqnM1.4}
\end{equation}

The statistical ensemble average of an observable operator $A$ at time $t$ ($t > t_i$) is defined as $\langle A \rangle (t) = \text{Tr} [\rho_0 A_H(t)]$, where the observable operator in the Heisenberg picture is defined by $A_H(t)=U_H^{\dag}(t,t_i) A(t) U_H(t, t_i)$. It can be easily shown that, in the interaction picture,  
\begin{equation}
\langle A \rangle (t) = \text{Tr} [\rho_I (t)  A_I(t)] = \text{Tr} [\rho_0 S(t_i, t) A_I(t) S(t, t_i)] , \label{eqnM1.5}
\end{equation}
where $\rho_I(t) = S(t, t_i) \rho_0 S(t_i, t)$ is the density matrix in the interaction picture, and $A_I(t)$ is defined in the same way as $V_I(t)$ in Eq. (\ref{eqnM1.4}). A perturbation detection theory for the observable operator $A$ can be established by the perturbation expansions of the $S$-matrix as 
\begin{eqnarray}
S(t,t_i) = \sum_{n=0}^{+\infty} \frac{1}{n!} \left(-\frac{i}{\hbar}\right)^n \int_{t_i}^{t} d t_n \cdots \int_{t_i}^{t} d t_1 T_t [V_I(t_n)\cdots V_I(t_1)] , \qquad \label{eqnM1.6}
\end{eqnarray}
and $S(t_i, t) = [S(t, t_i)]^{\dag}$. This is a general $S$-matrix perturbation detection theory for the detection interaction $V$ relevant system. 

\subsection{$S$-matrix perturbation theory for ARPES} \label{seca1.2}

The combined system for the ARPES measurement includes the target electrons, the incident photons and the emitted photoelectrons. Before each photon-pulse detection, the combined system has a Hamiltonian $H_{A,0}$. At the beginning time $t_i$ of each photon-pulse detection, the electron-photon interaction $V_A$ is turned on. The relevant $S$-matrix is defined by $S_A(t_f, t_i) = T_t \exp [-\frac{i}{\hbar}\int_{t_i}^{t_f} d t V_{A,I}(t)]$.  Let us introduce the time-window function $F(t)=\theta(t+ \Delta t_d/2) - \theta(t-\Delta t_d/2)$ for each photon-pule detection, where $t_i = -\Delta t_d/2$ and $t_f = + \Delta t_d/2$. The $S$-matrix can be expressed into another form as given in Section \ref{sec2}, $S_A=T_t \exp [-\frac{i}{\hbar} \int_{-\infty}^{+\infty} dt \, V_{A,I}(t) \cdot F(t)]$. Because $\Delta t_d \gg t_c, \Delta t_p$, we have set $t_i \rightarrow - \infty$ and $t_f \rightarrow + \infty$ in the final derivations.

The initial states of the combined system at the beginning time $t_i$ of each photon-pulse detection can be described by the density matrix $\rho_{A} = \sum_I P_{A,I} |\Phi_{A,I}^{(1)}\rangle \langle \Phi_{A,I}^{(1)} |$, where the distribution function $P_{A,I}$ is defined by $P_{A,I} = \frac{1}{Z} e^{-\beta E_\alpha} P_A(\mathbf{q},\lambda)$ and $\sum_I=\sum_{\alpha \mathbf{q}\lambda\chi_i}$. $Z = \text{Tr}(e^{-\beta H_s})$, where $H_s$ is the target-electron Hamiltonian with eigenvalues $E_\alpha$. Since the photoemission probability of the ARPES measurement is mainly dominated by the single-photoelectric processes, it can be defined by 
\begin{equation}
\widebar{\Gamma}_A^{(1)} = \text{Tr} [\rho_{A,I}^{(1)}(t_f) 1_A^{(1)}] = \text{Tr} [\rho_A S_A^{(1)}(t_i,t_f) 1_A^{(1)} S_A^{(1)}(t_f, t_i)] . \label{eqnM2.1}
\end{equation}    
Here $\rho_{A,I}^{(1)}(t_f)$ is the first-order part of the density matrix $\rho_{A,I}(t_f) = S_A(t_f, t_i) \rho_A S_A(t_i,t_f)$. $S_A^{(1)}(t_f,t_i)$ is the first-order perturbation expansion of the $S_A$-matrix and defined as $S_A^{(1)} (t_f,t_i)=-\frac{i}{\hbar} \int_{t_i}^{t_f} d t_1 V_{A,I}(t_1)$, and $S_A^{(1)} (t_i,t_f) = S_A^{(1)\dag}(t_f,t_i)$. In Eq. (\ref{eqnM2.1}), $1_A^{(1)}$ is a projection operator for the final states of the single-photoelectric processes of the ARPES measurement and defined by $1_A^{(1)} = \sum_{F} | \Phi^{(1)}_{A,F} \rangle \langle \Phi^{(1)}_{A,F} |$ with $\sum_{F}=\sum_{\beta\mathbf{q\lambda}\chi_f n^{(d)}}$. $\widebar{\Gamma}_A^{(1)}$ can be reexpressed into the following form as 
\begin{equation}
\widebar{\Gamma}_A^{(1)} = \sum_I P_{A,I} \langle \Phi_{A,I}^{(1)} (t_f)| 1_A^{(1)} | \Phi_{A,I}^{(1)} (t_f)  \rangle , \label{eqnM2.3}
\end{equation}  
where $| \Phi_{A,I}^{(1)} (t_f)  \rangle = S_A^{(1)}(t_f, t_i) | \Phi_{A,I}^{(1)} \rangle$. The photoemission probability of the ARPES measurement can be shown to follow  
\begin{equation}
\widebar{\Gamma}_A^{(1)}=\frac{1}{Z}\sum_{I F} e^{-\beta E_\alpha} P_A(\mathbf{q},\lambda) \big| \langle \Phi^{(1)}_{A,F} | S_A^{(1)} (t_f, t_i) | \Phi^{(1)}_{A,I} \rangle \big|^2 , \label{eqnM2.4} 
\end{equation}
where $\sum_{IF}=\sum_{\alpha\beta\mathbf{q\lambda}\chi_i\chi_f n^{(d)}}$. This is one main result of the $S$-matrix perturbation theory for the ARPES. Following the detailed derivation in Section II of \textbf{Supplementary material}, we can obtain the results for the ARPES in Section \ref{sec2}.

\subsection{$S$-matrix perturbation theory for post-experiment cARPES} \label{seca1.3}

From the above discussion on the $S$-matrix perturbation theory for the ARPES, the coincidence probability of {\it pulse}-resolved two photoelectric processes for the {\it post-experiment} cARPES measurement can be defined by  
\begin{equation}
\widebar{\Gamma}_A^{(2)} = \text{Tr} [\rho_{A,I}^{(2)}(t_f) 1_A^{(2)}] = \text{Tr} [\rho_A S_A^{(2)}(t_i,t_f) 1_A^{(2)} S_A^{(2)}(t_f, t_i) ] , \label{eqnM3.1}
\end{equation}     
where $\rho_{A,I}^{(2)}(t_f)$ is the second-order part of the density matrix $\rho_{A,I}(t_f)$. $S_A^{(2)}(t_f,t_i)$ is the second-order perturbation expansion of the $S_A$-matrix and defined by $S_A^{(2)}(t_f, t_i) = \frac{1}{2}(-\frac{i}{\hbar})^2 \iint_{t_i}^{t_f} d t_2 d t_1 T_t [V_{A,I}(t_2) V_{A,I}(t_1)]$, and $S_A^{(2)} (t_i,t_f) = S_A^{(2)\dag}(t_f,t_i)$. $1_A^{(2)}$ is a projection operator for the final states of the cARPES measurement and defined as $1_A^{(2)} = \sum_{F} | \Phi^{(2)}_{A,F} \rangle \langle \Phi^{(2)}_{A,F} |$ with $\sum_{F}=\sum_{\beta\mathbf{q\lambda}\chi_f n^{(d)}}$. Similarly, $\widebar{\Gamma}_A^{(2)}$ can be reexpressed into the below form as 
\begin{equation}
\widebar{\Gamma}_A^{(2)} = \sum_I P_{A,I} \langle \Phi_{A,I}^{(2)} (t_f)| 1_A^{(2)} | \Phi_{A,I}^{(2)} (t_f)  \rangle , \label{eqnM3.2}
\end{equation}  
where $| \Phi_{A,I}^{(2)} (t_f)  \rangle = S_A^{(2)}(t_f, t_i) | \Phi_{A,I}^{(2)} \rangle$. The coincidence probability of {\it pulse}-resolved two photoelectric processes for the {\it post-experiment} cARPES measurement can be shown to follow
\begin{equation}
\widebar{\Gamma}_A^{(2)}=\frac{1}{Z}\sum_{I F} e^{-\beta E_\alpha} P_A(\mathbf{q},\lambda) \big| \langle \Phi^{(2)}_{A,F} | S_A^{(2)} (t_f, t_i) | \Phi^{(2)}_{A,I} \rangle \big|^2 , \label{eqnM3.3} 
\end{equation}
where $\sum_{IF}=\sum_{\alpha\beta\mathbf{q\lambda}\chi_i\chi_f n^{(d)}}$. This is one main result of the $S$-matrix perturbation theory for the {\it post-experiment} cARPES. From the detailed derivation in Section III of \textbf{Supplementary material}, we can obtain the results for the {\it post-experiment} cARPES in Section \ref{sec2}.  

One more interesting result is given as follows. Let us introduce a pair-photoelectron operator for the {\it post-experiment} cARPES coincidence detection, $J_{\mathbf{k}_1\sigma_1 \mathbf{k}_2\sigma_2} = d^\dag_{\mathbf{k}_2\sigma_2} d_{\mathbf{k}_2\sigma_2} d^\dag_{\mathbf{k}_1\sigma_1} d_{\mathbf{k}_1\sigma_1}$, where $d^\dag_{\mathbf{k}_1\sigma_1} \, (d_{\mathbf{k}_1\sigma_1})$ and $d^\dag_{\mathbf{k}_2\sigma_2} \, (d_{\mathbf{k}_2\sigma_2})$ are the creation (annihilation) operators of the photoelectrons arrived at two detectors $D_1$ and $D_2$, respectively. From Eq. (\ref{eqnM1.5}), the statistical observation value of $J_{\mathbf{k}_1\sigma_1 \mathbf{k}_2\sigma_2}$ at the observation time $t_f$ can be defined by
\begin{equation}
\langle J_{\mathbf{k}_1\sigma_1 \mathbf{k}_2\sigma_2}\rangle = \text{Tr}[\rho_A S_A(t_i, t_f) J_{I,\mathbf{k}_1\sigma_1 \mathbf{k}_2\sigma_2} (t_f) S_A(t_f, t_i)] , \label{equM3.4}
\end{equation}
where $J_{I,\mathbf{k}_1\sigma_1 \mathbf{k}_2\sigma_2} (t)$ is the pair-photoelectron operator in the interaction picture. From the discussion in Section III of \textbf{Supplementary material}, we can show the following relation 
\begin{equation}
\widebar{\Gamma}_A^{(2)} \simeq \langle J_{\mathbf{k}_1\sigma_1 \mathbf{k}_2\sigma_2}\rangle .  \label{eqnM3.5}  
\end{equation}  
This is a very interesting result that the coincidence probability $\widebar{\Gamma}_A^{(2)}$ we have introduced for the cARPES measurement is equivalent approximately to the observation value of the pair-photoelectron operator $ J_{\mathbf{k}_1\sigma_1 \mathbf{k}_2\sigma_2}$, the latter of which is closely related to a pair-photoelectron current operator introduced in the reference \citep{DevereauxPRB2023}. 

With a similar derivation, we can establish the $S$-matrix perturbation theories for the INS and the {\it post-experiment} cINS. More detailed informations can be found in Section IV and V of \textbf{Supplementary material}.






\begin{thebibliography}{24}
\bibitem [{rem({\natexlab{a}})}]{remark1}%
  \BibitemOpen
  \bibinfo {note} {In other case such as $a_1(n)=2$ and $a_2(n)=1$, three
  photoelectrons are detected coincidently and the coincidence counting is
  defined for the coincidence probability of three photoelectric processes.
  This coincidence probability is relevant to a three-body Bethe-Salpeter wave
  function, $\Phi^{(3)}_{A,\alpha\beta}(\mathbf{k}_1 \sigma_1 t_1; \mathbf{k}_2
  \sigma_2 t_2; \mathbf{k}_3 \sigma_3 t_3) = \langle \Psi_{\beta} | T_t
  c_{\mathbf{k}_3\sigma_3}(t_3) c_{\mathbf{k}_2\sigma_2}(t_2)
  c_{\mathbf{k}_1\sigma_1} (t_1) | \Psi_{\alpha }\rangle$. All data like this
  case should be deleted during data analyses, or they would lead to
  experimental errors for the coincidence probability of two photoelectric
  processes.}\BibitemShut {Stop}%
\bibitem [{rem({\natexlab{b}})}]{remark2}%
  \BibitemOpen
  \bibinfo {note} {In order that the coincidence detection from each incident
  pulse is completely independent, it also requires $\Delta t_d \gg t_T$, where
  $t_T$ is the characteristic time scale for the target electrons to return to
  the initial thermodynamic equilibrium state.}\BibitemShut {Stop}%
\end{thebibliography}

%

\end{document}


\title{ Supplementary material for ``Post-experiment coincidence detection techniques for direct detection of two-body correlations" }

\author{Dezhong Cao}
\affiliation{ Department of Physics, Yantai University, Yantai 264005, People's Republic of China }

\author{Yuehua Su}
\email{suyh@ytu.edu.cn}
\affiliation{ Department of Physics, Yantai University, Yantai 264005, People's Republic of China }

\maketitle

\tableofcontents 

\section{A general $S$-matrix perturbation detection theory} \label{SecA1}

Let us consider a system with a time-independent Hamiltonian $H_0$ at time $t \leq t_i$. At time $t_i$, a detection interaction $V(t)$ is turned on and the Hamiltonian of the system becomes into the following form as  
\begin{equation}
H(t) = H_0(t) + V(t) , \label{eqnA1.1}
\end{equation}  
where $H_0(t)$ may become time dependent after time $t_i$. Suppose the system is in a quantum state $|\Psi(t_i)\rangle$ at time $t_i$. After time $t_i$, this quantum state follows the Schr\"odinger equation $i\hbar \frac{\partial}{\partial t} |\Psi_S(t)\rangle = H(t) |\Psi_S(t)\rangle$ and thus follows a time evolution as 
\begin{equation}
|\Psi_S(t)\rangle = U_H(t, t_i) |\Psi(t_i)\rangle, \,\, U_H(t, t_i) = T_t \exp[-\frac{i}{\hbar}\int_{t_i}^{t} d t_1 H(t_1)] ,  \label{eqnA1.2} 
\end{equation}
where $T_t$ is a time-ordering operator. Proof of this result is given as follows. If the time evolution of the quantum state $|\Psi_S(t)\rangle$ follows Eq. (\ref{eqnA1.2}), then 
\begin{eqnarray}
\frac{\partial }{\partial t} |\Psi_S(t)\rangle 
&=& \lim_{\Delta t \rightarrow 0} \frac{1}{\Delta t} [ e^{-\frac{i}{\hbar} H(t) \Delta t} -1 ] T_t \exp [-\frac{i}{\hbar}\int_{t_i}^{t} d t_1 H(t_1)] |\Psi(t_i)\rangle) \notag \\
&=& -\frac{i}{\hbar} H(t)|\Psi_S(t)\rangle) . \label{eqnA1.3}
\end{eqnarray}
Therefore, $|\Psi_S(t)\rangle$ follows the Schr\"odinger equation. Eq. (\ref{eqnA1.2}) describes the time evolution of the quantum state in the Schr\"odinger picture. 

Consider an observable operator $A$. The expectation value of $A$ at time $t$ in the quantum state $\Psi_S(t)$ is given by 
\begin{equation}
\langle A \rangle (t) = \langle \Psi_S(t) | A(t) | \Psi_S(t) \rangle = \langle \Psi(t_i)  | U_H^\dag (t,t_i) A(t) U_H(t,t_i) | \Psi(t_i) \rangle . \label{eqnA1.4} 
\end{equation}
Introduce the representation of $A$ in the Heisenberg picture,  
\begin{equation}
A_H(t) = U_H^\dag (t,t_i) A(t) U_H(t,t_i) , \label{eqnA1.5}
\end{equation}
the expectation value of $A$ at time $t$ can be reexpressed in the Heisenberg picture as 
\begin{equation}
\langle A \rangle (t) =  \langle \Psi(t_i)  | A_H(t) | \Psi(t_i) \rangle. \label{eqnA1.6}
\end{equation}
Introduce the representation of $A$ in the interaction picture,  
\begin{equation}
A_I(t) = U_0^\dag (t,t_i) A(t) U_0(t,t_i), \,\, U_0(t,t_i) = T_t \exp [-\frac{i}{\hbar}\int_{t_i}^{t} d t_1 H_0(t_1)] . \label{eqnA1.7}
\end{equation}  
The expectation value of $A$ at time $t$ can be expressed in the interaction picture as 
\begin{equation}
\langle A \rangle (t) =  \langle \Psi_I(t)  | A_I(t) | \Psi_I(t) \rangle, \label{eqnA1.8}
\end{equation}
where the quantum state in the interaction picture $| \Psi_I(t) \rangle$ follows 
\begin{equation}
| \Psi_I(t) \rangle = S(t,t_i) | \Psi(t_i) \rangle, \,\, S(t,t_i) = U_0^\dag (t,t_i) U_H(t,t_i) .  \label{eqnA1.9} 
\end{equation}
Let us now consider the time evolution of the $S$-matrix. When $t > t_i$, $U_0^\dag (t,t_i)$ follows  
\begin{equation}
U_0^\dag (t,t_i) = [U_0(t,t_i)]^\dag = \widetilde{T}_t \exp [-\frac{i}{\hbar} \int_{t}^{t_i} d t_1 H_0(t_1)] , \label{eqnA1.10}
\end{equation}
where $\widetilde{T}_t$ is an anti-chronological time-ordering operator.  It can be shown that   
\begin{eqnarray}
\frac{\partial}{\partial t} U_0^\dag(t,t_i) &=& \lim_{\Delta t \rightarrow 0} \frac{1}{\Delta t} [U_0^\dag(t+\Delta t, t_i) - U_0^\dag(t, t_i) ] \notag \\
&=& \lim_{\Delta t \rightarrow 0} \frac{1}{\Delta t} \widetilde{T}_t \exp [-\frac{i}{\hbar} \int_{t}^{t_i} d t_1 H_0(t_1) ] [ e^{\frac{i}{\hbar} H_0(t) \Delta t} - 1] \notag \\
&=& U_0^\dag(t,t_i) [\frac{i}{\hbar} H_0(t)] . \label{eqnA1.11} 
\end{eqnarray}
From Eq. (\ref{eqnA1.3}), $U_H(t,t_i)$ can be shown to follow
\begin{equation}
\frac{\partial}{\partial t} U_H(t,t_i) = -\frac{i}{\hbar} H(t) U_H(t, t_i) . \label{eqnA1.12}
\end{equation}
Thus, $S(t, t_i)$ follows  
\begin{eqnarray}
\frac{\partial}{\partial t} S(t, t_i) &=& \left[\frac{\partial}{\partial t} U_0^\dag(t,t_i) \right] U_H(t, t_i) + U_0^\dag (t,t_i) \left[\frac{\partial}{\partial t} U_H(t,t_i) \right]  \notag \\
&=& U_0^\dag (t, t_i) [\frac{i}{\hbar} H_0(t) - \frac{i}{\hbar}H(t)] U_H(t, t_i) \notag \\
&=& -\frac{i}{\hbar} V_I(t) S(t, t_i) . \label{eqnA1.13}
\end{eqnarray}
Here $V_I(t)$ is the representation of $V(t)$ in the interaction picture defined in the same way as $A_I(t)$ in Eq. (\ref{eqnA1.7}). The solution of the $S$-matrix equation, Eq. (\ref{eqnA1.13}), can be shown to follow 
\begin{equation}
S(t,t_i)= T_t \exp [-\frac{i}{\hbar} \int_{t_i}^{t} d t_1 V_I(t_1)] = \sum_{n=0}^{+\infty} \frac{1}{n!} \left(-\frac{i}{\hbar}\right)^n \int_{t_i}^{t} d t_n \cdots \int_{t_i}^{t} d t_1 T_t [V_I(t_n)\cdots V_I(t_1)]. \label{eqnA1.14}
\end{equation}

Suppose at time $t\leq t_i$, the system is in thermodynamic equilibrium which is represented by an ensemble with a density matrix 
\begin{equation}
\rho_0 = \frac{1}{Z} e^{-\beta H_0} = \frac{1}{Z} \sum_\alpha e^{-\beta E_\alpha} | \Psi_\alpha \rangle \langle \Psi_\alpha | , \label{eqnA1.15}
\end{equation} 
where the partition function $Z = \text{Tr}(e^{-\beta H_0})$ and $|\Psi_\alpha \rangle$ are the eigenstates of $H_0$ with the corresponding eigenvalues $E_\alpha$. The statistical ensemble average of the observable operator $A$ at time $t$ ($t > t_i$) can be described in three different pictures, the Heisenberg picture, the Schr\"odinger picture and the interaction picture as following:  
\begin{equation}
\langle A \rangle (t) = \text{Tr} [\rho_0 A_H(t)] 
= \text{Tr} [\rho_S(t) A(t)] 
= \text{Tr} [\rho_I(t) A_I(t)] . \label{eqnA1.16}
\end{equation}
Here the density matrix at time $t$ in the Schr\"odinger picture follows
\begin{equation}
\rho_S(t) = \frac{1}{Z} \sum_\alpha e^{-\beta E_\alpha} | \Psi_{\alpha,S} (t) \rangle \langle \Psi_{\alpha,S}(t) | = U_H(t,t_i) \rho_0 U_H^\dag(t,t_i) , \label{eqnA1.17}
\end{equation}
and the density matrix in the interaction picture follows
\begin{equation}
\rho_I(t) = \frac{1}{Z} \sum_\alpha e^{-\beta E_\alpha} | \Psi_{\alpha,I} (t) \rangle \langle \Psi_{\alpha,I}(t) | =  S(t,t_i) \rho_0 S(t_i,t) , \label{eqnA1.18}
\end{equation}
where $S(t_i,t)= [S(t,t_i)]^\dag$ is given by 
\begin{equation}
S(t_i,t)= \widetilde{T}_t \exp [-\frac{i}{\hbar} \int_{t}^{t_i} d t_1 V_I^\dag (t_1)] = \sum_{n=0}^{+\infty} \frac{1}{n!} \left(-\frac{i}{\hbar}\right)^n \int_{t}^{t_i} d t_1 \cdots \int_{t}^{t_i} d t_n \widetilde{T}_t [V_I^\dag(t_1)\cdots V_I^\dag (t_n)]. \label{eqnA1.19}
\end{equation}

The $S$-matrix perturbation detection theory can be established in the interaction picture by the perturbation expansions of the $S$-matrices as Eqs. (\ref{eqnA1.14}) and (\ref{eqnA1.19}). For example, the statistical ensemble average of the observable operator $A$ is defined by 
\begin{equation}
\langle A \rangle (t) = \text{Tr} [\rho_I(t) A_I(t)] =\text{Tr} [\rho_0 S(t_i ,t) A_I(t) S(t, t_i)] . \label{eqnA1.20}  
\end{equation}
When the $S$-matrices are expanded to the first-order perturbations, we can obtain the famous Kubo formula from the following expression
\begin{equation}
\langle A \rangle (t) = \langle A_I(t) \rangle_0 - \frac{i}{\hbar} \int_{t_i}^{t} d t_1 \langle [A_I(t), V_I(t_1)] \rangle_0 , \label{eqnA1.21} 
\end{equation} 
which can lead us the linear response function of the system to the external perturbation interaction $V$. Here $\langle A \rangle_0 = \text{Tr}(\rho_0 A)$, and $V^\dag(t) = V(t)$ is assumed. 
It should be noted that from Eq. (\ref{eqnA1.20}) and the perturbation expansions of $S(t, t_i)$ and $S(t_i, t)$, it can be easily shown that, when $H_0(t)$ is time independent for $t>t_i$, all operators in the interaction picture can be equivalently defined by 
\begin{equation}
A_I(t) = e^{\frac{i}{\hbar} H_0 t} A(t) e^{-\frac{i}{\hbar} H_0 t} , \label{eqnA1.22}
\end{equation}
which is the initial time $t_i$ independent. 

Let us consider the time-resolved angle-resolved photoemission spectroscopy (TR-ARPES) \citep{FreericksTRARPES2009}. At time $t\leq t_i$, $H_0=H_s$ is the Hamiltonian of the target electrons. At time $t_i$, the pump field and the probe field are turned on and $H_0(t)=H_s + H_{pump}(t)$ and $V(t)=H_{prob}(t)$, where $H_{pump}(t)$ and $H_{prob}(t)$ define the Hamiltonians and the interactions of the pump field and the probe field to the target electrons, respectively. Consider the statistical ensemble average of a photoelectron current operator $J_d$ at time $t > t_i$. Since the finite observation value of $J_d$ in the TR-ARPES is mainly dominated by the single-photoelectric processes, the observation value of $J_d$ can be calculated approximately by 
\begin{equation}
\langle J_d \rangle (t) = \text{Tr} [\rho_0 S_1(t_i ,t) J_{d,I}(t) S_1(t, t_i)] , \label{eqnA1.23}
\end{equation}   
where $S_1(t,t_i)$ and $S_1(t_i, t)$ are the first-order perturbation expansions of the $S$-matrices $S(t,t_i)$ and $S(t_i,t)$, respectively, and follow
\begin{eqnarray}
S_1(t,t_i) &=& -\frac{i}{\hbar} \int_{t_i}^t d t_1 V_I(t_1) , \label{eqnA1.24-1} \\
S_1(t_i,t) &=& + \frac{i}{\hbar} \int_{t_i}^t d t_1 V_I^\dag(t_1) . \label{eqnA1.24-2}  
\end{eqnarray}
Therefore, the observation value of $J_d$ in the TR-ARPES follows 
\begin{equation}
\langle J_d \rangle (t) = \frac{1}{\hbar^2} \iint_{t_i}^t d t_2 d t_1 \langle V_I^\dag (t_2) J_{d,I}(t) V_I(t_1)  \rangle_0 . \label{eqnA1.25}
\end{equation}
This is one main result for the TR-ARPES which has been obtained previously by using the non-equilibrium Green's function theory \citep{FreericksTRARPES2009}. It shows that the $S$-matrix perturbation detection theory we have developed here is equivalent to the non-equilibrium Green's function theory in description of the non-equilibrium dynamical physics of the target matter. Without the pump field, i.e., $H_{pump}(t) = 0$, Eq. (\ref{eqnA1.25}) can recover the previous results for the conventional ARPES \citep{DevereauxPRB2023}.

\section{$S$-matrix perturbation theory for ARPES}  \label{SecA2} 

The combined system for the angle-resolved photoemission spectroscopy (ARPES) measurement includes the target electrons with a Hamiltonian $H_s$, the incident photons with a Hamiltonian $H_p = \sum_{\mathbf{q}\lambda} \hbar \omega_{\mathbf{q}} (a^\dag_{\mathbf{q}\lambda} a_{\mathbf{q}\lambda}+\frac{1}{2})$, and the emitted photoelectrons with a Hamiltonian $H_d = \sum_{\mathbf{k}\sigma} \varepsilon_{\mathbf{k}}^{(d)} d^\dag_{\mathbf{k}\sigma} d_{\mathbf{k}\sigma}$. Here $a^\dag_{\mathbf{q}\lambda}$ and $a_{\mathbf{q}\lambda}$ are the creation and annihilation operators for the photons with momentum $\mathbf{q}$ and polarization $\lambda$, and $d_{\mathbf{k}\sigma}^{\dag}$ and $d_{\mathbf{k}\sigma}$ are the creation and annihilation operators for the photoelectrons with momentum $\mathbf{k}$ and spin $\sigma$. The Hamiltonian for the ARPES measurement is given by 
\begin{equation}
H_A = H_{A,0} + V_A , \label{eqnA2.1}
\end{equation}
where $H_{A,0} = H_s + H_p + H_d$, and the electron-photon interaction $V_A$ is defined by
\begin{equation}
V_A=\sum_{\mathbf{k}\sigma\mathbf{q}\lambda} g_A(\mathbf{k};\mathbf{q},\lambda) d^\dag_{\mathbf{k+q}\sigma}c_{\mathbf{k}\sigma} a_{\mathbf{q}\lambda} , \label{eqnA2.2}
\end{equation}
where $c_{\mathbf{k}\sigma}$ is the annihilation operator for the target electrons with momentum $\mathbf{k}$ and spin $\sigma$. Here $V_A$ only involves the photon-absorption and photoelectron-emission processes for the ARPES measurement. The relevant $S$-matrix for the ARPES measurement is defined by 
\begin{equation}
S_A(t_f, t_i) = T_t \exp [-\frac{i}{\hbar}\int_{t_i}^{t_f} d t V_{A,I}(t)] , \,\, V_{A,I}(t)=e^{\frac{i}{\hbar} H_{A,0} t } V_A e^{-\frac{i}{\hbar} H_{A,0} t} . \label{eqnA2.3}
\end{equation} 
For the photon-pulse source, $t_i$ and $t_f$ define the time window for every one photon-pulse detection, which can be described by a time-window function $F(t)=\theta(t+ \frac{1}{2} \Delta t_d) - \theta(t-\frac{1}{2} \Delta t_d)$, where $\theta$ is the step function and $\Delta t_d$ is the time window between sequential two photon pulses. From the definition of $F(t)$, it shows that $t_i = -\frac{1}{2}\Delta t_d$ and $t_f = + \frac{1}{2}\Delta t_d$. Because $\Delta t_d \gg t_c, \Delta t_p$, where $t_c$ is the characteristic time scale of the physics we are interested in and $\Delta t_p$ is the time width of the photon pulse, we will set $t_i \rightarrow -\infty$ and $t_f \rightarrow +\infty$ in the final derivations.  
 
Let us consider the single-photoelectric processes for the ARPES measurement, where the incident photons from the photon pulses are in the initial states $|\chi_i(\mathbf{q} \lambda) \rangle$ with a distribution function $P_A(\mathbf{q},\lambda)$ and the emitted photoelectrons are focused with fixed momentum $\mathbf{k}$ and spin $\sigma$. 
The density matrix for the initial states of the combined system at the beginning time $t_i$ of every photon-pulse emission is defined by
\begin{equation}
\rho_A = \sum_I P_{A,I} |\Phi_{A,I}^{(1)}\rangle \langle \Phi_{A,I}^{(1)} |, \label{eqnA2.4}
\end{equation}
where the distribution function $P_{A,I}$ and the initial states $|\Phi_{A,I}^{(1)}\rangle$ are defined as
\begin{eqnarray}
&& P_{A,I} = \frac{1}{Z} e^{-\beta E_\alpha} P_A(\mathbf{q},\lambda), \label{eqnA2.5-1} \\
&& |\Phi_{A,I}^{(1)}\rangle = |\Psi_\alpha\rangle \otimes |\chi_{i}(\mathbf{q}\lambda) \rangle \otimes |0^{(d)}\rangle . \label{eqnA2.5-2}
\end{eqnarray}
Here $\sum_I=\sum_{\alpha \mathbf{q}\lambda\chi_i}$, $Z=\text{Tr}(e^{-\beta H_s})$ and $|\Psi_\alpha\rangle$ are the eigenstates of $H_s$ with the corresponding eigenvalues $E_{\alpha}$, $|0^{(d)}\rangle$ defines an initial photoelectron vacuum state. 
Since the photoemission probability of the ARPES measurement is mainly dominated by the single-photoelectric processes,  the total photoemission probability of all the single-photoelectric processes at the detection time $t_f$ can be defined by 
\begin{equation}
\widebar{\Gamma}_A^{(1,t)} = \text{Tr} [\rho_{A,I}^{(1)}(t_f)] = \text{Tr} [S_A^{(1)}(t_f, t_i) \rho_A S_A^{(1)}(t_i,t_f) ] , \label{eqnA2.6}
\end{equation}     
where $\rho_{A,I}^{(1)}(t_f)$ is the first-order part of the density matrix $\rho_{A,I}(t_f)=S_A(t_f, t_i)\rho_A S_A(t_i, t_f)$. $S_A^{(1)}(t_f,t_i)$ is the first-order perturbation expansion of the $S_A$-matrix, which is relevant to the single-photoelectric processes and defined as $S_A^{(1)} (t_f,t_i)=-\frac{i}{\hbar} \int_{t_i}^{t_f} d t V_{A,I}(t)$, and $S_A^{(1)} (t_i,t_f) = S_A^{(1)\dag}(t_f,t_i)$. Physically, $\widebar{\Gamma}_A^{(1,t)}$ can be expressed into the below form as
\begin{equation}
\widebar{\Gamma}_A^{(1,t)} = \sum_{I} P_{A,I} \langle \Phi^{(1)}_{A,I} (t_f) | \Phi^{(1)}_{A,I} (t_f) \rangle , \, \, |\Phi^{(1)}_{A,I} (t_f)\rangle = S_A^{(1)} (t_f, t_i) | \Phi^{(1)}_{A,I} \rangle . \label{eqnA2.7} 
\end{equation}
This clearly shows that $\widebar{\Gamma}_A^{(1,t)}$ defines the state probability of the combined system at time $t_f$ after the occurrence of all the single-photoelectric processes. Let us introduce a projection operator for the final states of the single-photoelectric processes with fixed photoelectron momentum $\mathbf{k}$ and spin $\sigma$,   
\begin{equation}
1_A^{(1)} = \sum_{F} | \Phi^{(1)}_{A,F} \rangle \langle \Phi^{(1)}_{A,F} |, \,\, |\Phi^{(1)}_{A,F}\rangle = |\Psi_\beta \rangle \otimes |\chi_f(\mathbf{q}\lambda)\rangle \otimes |n_{\mathbf{k}\sigma}^{(d)}\rangle , \label{eqnA2.8}
\end{equation}
where $\sum_F = \sum_{\beta\mathbf{q}\lambda \chi_f n^{(d)}}$, $|\Psi_\beta \rangle$ are the eigenstates of the target electrons, $|\chi_f(\mathbf{q}\lambda)\rangle$ describe the final photon states and $n_{\mathbf{k}\sigma}^{(d)}=0,1$ is the number of the photoelectrons arrived at one single-photoelectron detector. The photoemission probability of the ARPES measurement obtained by one single-photoelectron detector with focused momentum $\mathbf{k}$ and spin $\sigma$ can be defined by 
\begin{equation}
\widebar{\Gamma}_A^{(1)} = \sum_{I} P_{A,I} \langle \Phi^{(1)}_{A,I} (t_f) | 1_A^{(1)} | \Phi^{(1)}_{A,I}(t_f) \rangle , \label{eqnA2.9-1}
\end{equation}
which has another physically equivalent form as
\begin{equation}
\widebar{\Gamma}_A^{(1)} = \text{Tr} [\rho_{A,I}^{(1)}(t_f) 1_A^{(1)}] . \label{eqnA2.9-2}
\end{equation}  
It can be shown that $\widebar{\Gamma}_A^{(1)}$ follows 
\begin{equation}
\widebar{\Gamma}_A^{(1)}=\frac{1}{Z}\sum_{I F} e^{-\beta E_\alpha} P_A(\mathbf{q},\lambda) \big| \langle \Phi^{(1)}_{A,F} | S_A^{(1)} (t_f, t_i) | \Phi^{(1)}_{A,I} \rangle \big|^2 , \label{eqnA2.10} 
\end{equation}
where $\sum_{IF}=\sum_{\alpha\beta\mathbf{q\lambda}\chi_i\chi_f n^{(d)}}$. Eq. (\ref{eqnA2.10}) is one main result of the $S$-matrix perturbation theory for the ARPES. 

Let us first consider the photoemission probability of one single-photoelectric process with one initial state $| \Phi^{(1)}_{A,I} \rangle$ and one final state $| \Phi^{(1)}_{A,F} \rangle$, which can be defined by 
\begin{equation}
\widebar{\Gamma}_{A,IF}^{(1)}= \big| \langle \Phi^{(1)}_{A,F} | S_A^{(1)} (t_f, t_i) | \Phi^{(1)}_{A,I} \rangle \big|^2 .  \label{eqnA2.11} 
\end{equation}
When we introduce the simplified representations for the initial and final states as $| \Phi^{(1)}_{A,I} \rangle = |\Psi_\alpha; \chi_{i}(\mathbf{q}\lambda); 0^{(d)}\rangle $ and $| \Phi^{(1)}_{A,F} \rangle = |\Psi_\beta; \chi_{f}(\mathbf{q}\lambda); n_{\mathbf{k}\sigma}^{(d)}\rangle$, it can be shown that 
\begin{eqnarray}
\widebar{\Gamma}_{A,IF}^{(1)} 
&=& \Big| (-\frac{i}{\hbar}) \int_{t_i}^{t_f} d t \langle n_{\mathbf{k}\sigma}^{(d)}; \chi_f(\mathbf{q}\lambda); \Psi_\beta| V_{A,I}(t) |\Psi_\alpha; \chi_{i}(\mathbf{q}\lambda); 0^{(d)}\rangle \Big|^2 \notag \\
&=& \Big| (-\frac{i}{\hbar}) \int_{t_i}^{t_f} d t  \sum_{\mathbf{k}^\prime\sigma^\prime\mathbf{q}^\prime\lambda^\prime} g_A(\mathbf{k}^\prime;\mathbf{q}^\prime,\lambda^\prime) \langle n_{\mathbf{k}\sigma}^{(d)}; \chi_f(\mathbf{q}\lambda); \Psi_\beta| d^\dag_{\mathbf{k}^\prime+\mathbf{q}^\prime\sigma^\prime}(t) c_{\mathbf{k}^\prime\sigma^\prime}(t) a_{\mathbf{q}^\prime\lambda^\prime}(t) |\Psi_\alpha; \chi_{i}(\mathbf{q}\lambda); 0^{(d)}\rangle \Big|^2 \notag \\
&=& \Big| (-\frac{i}{\hbar}) \int_{t_i}^{t_f} d t \sum_{\mathbf{k}^\prime\sigma^\prime} g_A(\mathbf{k}^\prime-\mathbf{q};\mathbf{q},\lambda) \langle \Psi_\beta| c_{\mathbf{k}^\prime-\mathbf{q} \sigma}(t) |\Psi_\alpha\rangle  \langle\chi_f(\mathbf{q}\lambda)| a_{\mathbf{q}\lambda} |\chi_{i}(\mathbf{q}\lambda) \rangle \langle n_{\mathbf{k}\sigma}^{(d)}| d^\dag_{\mathbf{k}^\prime\sigma^\prime}  |0^{(d)}\rangle e^{i [\varepsilon_{\mathbf{k}^\prime}^{(d)}/\hbar -\omega_{\mathbf{q}}] t} \Big|^2 . \qquad \label{eqnA2.12}
\end{eqnarray}
Here the sum over $\mathbf{k}^\prime$ and $\sigma^\prime$ describes all possible photoemissions in the single-photoelectric process with the emitted photoelectrons created by $d^\dag_{\mathbf{k}^\prime\sigma^\prime}$. Physically, $\widebar{\Gamma}_{A,IF}^{(1)}$ can be reexpressed into the following simplified form as
\begin{equation}
\widebar{\Gamma}_{A,IF}^{(1)} = \Big| \sum_{m} \langle \Phi^{(1)}_{A,F} | \Phi^{(1)}_{m} (t_f) \rangle \Big|^2 , \label{eqnA2.13} 
\end{equation} 
where $| \Phi^{(1)}_{m} (t_f) \rangle$ with $m \equiv \mathbf{k}^\prime\sigma^\prime$ is defined by 
\begin{equation}
| \Phi^{(1)}_{m} (t_f) \rangle =  (-\frac{i}{\hbar}) \int_{t_i}^{t_f} d t \,\, g_A(\mathbf{k}^\prime-\mathbf{q};\mathbf{q},\lambda) d^\dag_{\mathbf{k}^\prime\sigma^\prime}(t) c_{\mathbf{k}^\prime-\mathbf{q} \sigma^\prime}(t) a_{\mathbf{q} \lambda}(t) |\Phi^{(1)}_{A,I}\rangle . \label{eqnA2.14}
\end{equation}
We introduce $C_{F,m} = \langle \Phi^{(1)}_{A,F} | \Phi^{(1)}_{m} (t_f) \rangle$ to describe the probability of one quantum state $| \Phi^{(1)}_{m} (t_f) \rangle$ in the final state $|\Phi^{(1)}_{A,F}\rangle$. Thus, $\widebar{\Gamma}_{A,IF}^{(1)} = \big|\sum_{m} C_{F,m} \big|^2 = \sum_{m_1 m_2} C^{\ast}_{F,m_1}C_{F,m_2}$, which defines all probability of the linear superposition quantum state $\sum_m |\Phi^{(1)}_{m} (t_f) \rangle$  in the final state $|\Phi^{(1)}_{A,F}\rangle$. In the realistic ARPES measurement, when each single-photoelectron detector focuses on the emitted photoelectrons with fixed momentum $\mathbf{k}$ and spin $\sigma$, the final state relevant to one single-photoelectron detector has definite momentum $\mathbf{k}$ and spin $\sigma$. Thus, the photoemission probability detected by this detector is only relevant to one special quantum state with $m=\mathbf{k} \sigma$. Therefore, we have
\begin{equation}
\widebar{\Gamma}_{A,IF}^{(1)} = \big| C_{F,\mathbf{k}\sigma} \big|^2 \delta_{\mathbf{k} \mathbf{k}^\prime}\delta_{\sigma\sigma^\prime} = \big| \langle \Phi^{(1)}_{A,F} | \Phi^{(1)}_{\mathbf{k}\sigma} (t_f) \rangle \big|^2 \delta_{\mathbf{k} \mathbf{k}^\prime}\delta_{\sigma\sigma^\prime}. \label{eqnA2.15}
\end{equation}

The above discussion shows that $\widebar{\Gamma}_{A,IF}^{(1)}$ defines all probability of different $\mathbf{k}^\prime \sigma^\prime$ relevant photoemissions detected by one single-photoelectron detector with one focused final state $|\Phi^{(1)}_{A,F}\rangle$. Physically, in each ARPES detection with one focused final state $|\Phi^{(1)}_{A,F}\rangle$, only one $\mathbf{k}^\prime\sigma^\prime$ relevant photoemission occurs with only one $\mathbf{k}^\prime\sigma^\prime$-photoelectron emitted from this photoemission. During this ARPES detection, the single-photoelectron detector detects the number of the photoelectron $n_{\mathbf{k}\sigma}^{(d)}$. $n_{\mathbf{k}\sigma}^{(d)}=0$ when the emitted photoelectron with $\mathbf{k}^\prime \not= \mathbf{k}$ and $\sigma^\prime\not= \sigma$, and $n_{\mathbf{k}\sigma}^{(d)}=1$ when $\mathbf{k}^\prime  = \mathbf{k}$ and $\sigma^\prime = \sigma$. Therefore, we can introduce a photoelectron-state factor $I_{A,d}^{(1)}$ to define the photoelectron states of the single-photoelectron detector. It can be defined by  
\begin{equation}
I_{A,d}^{(1)} = \big| \langle n_{\mathbf{k}\sigma}^{(d)}| d^\dag_{\mathbf{k}\sigma}  |0^{(d)}\rangle \big|^2 . \label{eqnA2.16}  
\end{equation}
$I_{A,d}^{(1)} = 0$ when no photoelectron arrives at the detector with $n_{\mathbf{k}\sigma}^{(d)}=0$, and $I_{A,d}^{(1)} = 1$ when one photoelectron arrives at the detector with $n_{\mathbf{k}\sigma}^{(d)}=1$. Therefore, the photoelectron-state factor plays a role to record the number of the photoelectrons arrived at the single-photoelectron detector. This is a crucial trick we introduced for the detection of the single-photoelectron or single-neutron detector as well as for the {\it post-experiment} coincidence counting method we have presented in the main text.      

With the trick to account for the photoelectron states, $\widebar{\Gamma}_{A,IF}^{(1)}$ can be expressed into the form as  
\begin{equation}
\widebar{\Gamma}_{A,IF}^{(1)}  = \Gamma_{A,\alpha\beta}^{(1)} \cdot I_{A,\chi}^{(1)} \cdot I_{A,d}^{(1)}, \label{eqnA2.17}  
\end{equation} 
where $\Gamma_{A,\alpha\beta}^{(1)}$ is a target-electron form factor, $I_{A,\chi}^{(1)}$ is a photon-state factor  and $I_{A,d}^{(1)}$ is the photoelectron-state factor. $I_{A,\chi}^{(1)}$ is defined by  
\begin{equation}
I_{A,\chi}^{(1)} = \big| \langle\chi_f(\mathbf{q}\lambda)| a_{\mathbf{q}\lambda} |\chi_{i}(\mathbf{q}\lambda) \rangle \big|^2 . \label{eqnA2.18} 
\end{equation}
The target-electron form factor $\Gamma_{A,\alpha\beta}^{(1)}$ is defined and calculated as following:   
\begin{eqnarray}
\Gamma_{A,\alpha\beta}^{(1)} &=& \frac{|g_A|^2}{\hbar^2} \Big| \int_{t_i}^{t_f} d t \langle \Psi_\beta| c_{\mathbf{k-q}\sigma}(t) |\Psi_\alpha\rangle e^{i  [\varepsilon_{\mathbf{k}}^{(d)}/\hbar - \omega_{\mathbf{q}}] t} \Big|^2 \notag \\
&=& \frac{|g_A|^2}{\hbar^2} \Big| \int_{t_i}^{t_f} d t \langle \Psi_\beta| e^{i H_s t/\hbar} c_{\mathbf{k-q}\sigma} e^{-i H_s t/\hbar} |\Psi_\alpha\rangle e^{i [\varepsilon_{\mathbf{k}}^{(d)}/\hbar - \omega_{\mathbf{q}}] t} \Big|^2 \notag \\
&=& \frac{|g_A|^2}{\hbar^2} \big| \langle \Psi_\beta| c_{\mathbf{k-q}\sigma} |\Psi_\alpha\rangle \big|^2 \cdot \Big| \int_{t_i}^{t_f} d t e^{i [(E_\beta - E_\alpha + \varepsilon_{\mathbf{k}}^{(d)})/\hbar -\omega_{\mathbf{q}}] t} \Big|^2 \notag \\ 
&=& \frac{2\pi |g_A|^2 \Delta t_d}{\hbar} \big| \langle \Psi_\beta| c_{\mathbf{k-q}\sigma} |\Psi_\alpha\rangle \big|^2 \, \delta[E_\beta - E_\alpha + \varepsilon_{\mathbf{k}}^{(d)} -\hbar\omega_{\mathbf{q}}] ,  \label{eqnA2.19}
\end{eqnarray}
where $g_A = g_A (\mathbf{k-q}; \mathbf{q},\lambda)$. 
In the last step to derive Eq. (\ref{eqnA2.19}), we have set $t_i=-\frac{1}{2}\Delta t_d$ and $t_f=+\frac{1}{2}\Delta t_d$, and $\frac{\sin^2 (a x)}{x^2}=\pi a \delta(x)$ when $a\rightarrow +\infty$ has been used with the limit $\Delta t_d \rightarrow +\infty$.   

The statistical average of the photoemission probability of the ARPES measurement can be calculated from Eq. (\ref{eqnA2.10}), which follows
\begin{equation}
\widebar{\Gamma}_A^{(1)}=\frac{1}{Z}\sum_{I F} e^{-\beta E_\alpha} P_A(\mathbf{q},\lambda) \cdot \Gamma_{A,\alpha\beta}^{(1)} \cdot I_{A,\chi}^{(1)} \cdot I_{A,d}^{(1)} , \label{eqnA2.20} 
\end{equation}  
where $\sum_{IF}=\sum_{\alpha\beta\mathbf{q}\lambda \chi_i\chi_f n^{(d)}}$ and $\Gamma_{A,\alpha\beta}^{(1)}$ is given by Eq. (\ref{eqnA2.19}). Let us introduce the single-particle Green's function $G_\sigma(\mathbf{k},\tau) = -\langle T_\tau c_{\mathbf{k}\sigma}(\tau) c^\dag_{\mathbf{k}\sigma}(0) \rangle$, where $\tau$ is an imaginary time. The corresponding imaginary-frequency Fourier transformation is defined by $G_\sigma(\mathbf{k},i\omega_n) = \int_0^{\beta} d \tau G_\sigma(\mathbf{k},\tau) e^{i\omega_n \tau}$. The single-particle spectral function $A(\mathbf{k},\sigma; E) = - 2\, \text{Im} \, G_{\sigma} (\mathbf{k}, i\omega_n \rightarrow E + i \delta^{+})$ can be shown to follow 
\begin{equation}
A(\mathbf{k},\sigma; E) = \frac{2\pi}{Z}\sum_{\alpha\beta} (e^{-\beta E_\alpha} + e^{-\beta E_\beta}) \big| \langle \Psi_\beta| c_{\mathbf{k}\sigma} |\Psi_\alpha\rangle \big|^2 \, \delta(E+E_\beta - E_\alpha) . \label{eqnA2.21}
\end{equation}
$\widebar{\Gamma}_A^{(1)}$ can be expressed into the following form as 
\begin{equation}
\widebar{\Gamma}_A^{(1)}=\sum_{I F} P_A(\mathbf{q},\lambda) \cdot \Gamma_{A}^{(1)} \cdot I_{A,\chi}^{(1)} \cdot I_{A,d}^{(1)} , \label{eqnA2.22} 
\end{equation} 
where $\sum_{I F}= \sum_{\mathbf{q}\lambda \chi_i \chi_f n^{(d)}}$, and $\Gamma_{A}^{(1)} = \frac{1}{Z}\sum_{\alpha\beta} e^{-\beta E_\alpha} \Gamma_{A,\alpha\beta}^{(1)}$ is given by
\begin{equation}
\Gamma_{A}^{(1)} = \frac{|g_A|^2 \Delta t_d}{\hbar} A(\mathbf{k}-\mathbf{q}, \sigma ; E_A^{(1)}) \cdot n_F(E_A^{(1)}) . \label{eqnA2.23}
\end{equation}  
Here $E_A^{(1)}$ is the transferred energy in the single-photoelectric process and $n_F(E)$ is the Fermi-Dirac distribution function. $E_A^{(1)}$ is defined as $E_A^{(1)} = \varepsilon_{\mathbf{k}}^{(d)} + \Phi -\hbar \omega_\mathbf{q}$, where $\varepsilon_{\mathbf{k}}^{(d)}$ is the photoelectron energy and $\hbar \omega_\mathbf{q}$ is the photon energy. Here the work function $\Phi$ has been included in the definition of $E_A^{(1)}$. 
Consider a simple case where there is only one incident photon and the final state is focused with one photoelectron. In this case, $I_{A,\chi}^{(1)}$ and $I_{A,d}^{(1)}$ can be given by   
\begin{equation}
I_{A,\chi}^{(1)} = \big| \langle 0_{\mathbf{q}\lambda}| a_{\mathbf{q}\lambda} |1_{\mathbf{q}\lambda} \rangle \big|^2 =1 , \,\, I_{A,d}^{(1)} = \big| \langle 1_{\mathbf{k}\sigma}^{(d)}| d^\dag_{\mathbf{k}\sigma}  |0^{(d)}\rangle \big|^2 =1 . \label{eqnA2.24}
\end{equation}
Thus, we can recover the previous result of the ARPES \citep{SuZhang2020} that $\widebar{\Gamma}_A^{(1)}=\Gamma_{A}^{(1)}$. 

It should be remarked that the photoelectron-state factor $I_{A,d}^{(1)}$ makes us to obtain the {\it absolute} counting of the photoemission probability in realistic ARPES measurement, with zero counting when $n_{\mathbf{k}\sigma}^{(d)}=0$ and $I_{A,d}^{(1)}=0$ and finite counting when $n_{\mathbf{k}\sigma}^{(d)}=1$ and $I_{A,d}^{(1)}=1$. This is different from the conventional ARPES measurement, where only the signals with $n_{\mathbf{k}\sigma}^{(d)}=1$ and $I_{A,d}^{(1)}=1$ are recorded and only the {\it relative} photoemission probability can be obtained. The trick to introduce the photoelectron-state factors is crucial for the {\it post-experiment} coincidence detection techniques we have proposed in the main text.   

Let us consider a photoelectron operator $J_{\mathbf{k}\sigma} = d^\dag_{\mathbf{k}\sigma} d_{\mathbf{k}\sigma}$ for one single-photoelectron detector. The observation value of $J_{\mathbf{k}\sigma}$ at the observation time $t_f$ is defined by its statistical ensemble average following Eq. (\ref{eqnA1.20}), $\langle J_{\mathbf{k}\sigma}\rangle = \text{Tr}[\rho_A S_A(t_i, t_f) J_{I,\mathbf{k}\sigma} (t_f) S_A(t_f, t_i)]$. Since the observation value of $J_{\mathbf{k}\sigma}$ is mainly dominated by the single-photoelectric processes, it can be calculated approximately by 
\begin{eqnarray}
\langle J_{\mathbf{k}\sigma }\rangle 
&\simeq& \text{Tr}[\rho_A S_A^{(1)}(t_i, t_f) d^{\dag}_{\mathbf{k}\sigma}(t_f) d_{\mathbf{k}\sigma}(t_f) S_A^{(1)}(t_f, t_i)]  \notag \\
& = & \text{Tr}[\rho_A S_A^{(1)}(t_i, t_f) d^{\dag}_{\mathbf{k}\sigma}(t_f) \, 1_{A}^{\prime (1)} \, d_{\mathbf{k}\sigma}(t_f) S_A^{(1)}(t_f, t_i)]  \notag \\
&=& \sum_{I F^\prime} P_{A,I}  \langle \Phi^{(1)}_{A,I} | S_A^{(1)} (t_i, t_f) d^\dag_{\mathbf{k}\sigma} (t_f) | \Phi^{(1)}_{A,F^\prime} \rangle \langle \Phi^{(1)}_{A,F^{\prime}} | d_{\mathbf{k}\sigma} (t_f) S_A^{(1)} (t_f, t_i) | \Phi^{(1)}_{A,I} \rangle  \notag \\
&=& \sum_{I F^\prime} P_{A,I}  \big| \langle \Phi^{(1)}_{A,F^{\prime}} | d_{\mathbf{k}\sigma} S_A^{(1)} (t_f, t_i) | \Phi^{(1)}_{A,I} \rangle \big|^2 \notag \\
&=& \frac{1}{Z}\sum_{I F^\prime} e^{-\beta E_\alpha} P_A(\mathbf{q}\lambda) \big| \langle \Phi^{(1)}_{A,F} | S_A^{(1)} (t_f, t_i) | \Phi^{(1)}_{A,I} \rangle \big|^2 . \label{eqnA2.25}
\end{eqnarray} 
Here $1_{A}^{\prime (1)}=\sum_{F^{\prime}} |\Phi^{(1)}_{A,F^{\prime}}\rangle \langle \Phi^{(1)}_{A,F^{\prime}}|$ with $\sum_{F^\prime}=\sum_{\beta\mathbf{q}\lambda\chi_f\widebar{n}^{(d)}}$ and $|\Phi^{(1)}_{A,F^{\prime}}\rangle = |\Psi_\beta \rangle \otimes |\chi_f(\mathbf{q}\lambda)\rangle \otimes |\widebar{n}_{\mathbf{k}\sigma}^{(d)} \rangle$. The relation between $|\Phi^{(1)}_{A,F}\rangle$ and $|\Phi^{(1)}_{A,F^\prime}\rangle$ is defined by $|\Phi^{(1)}_{A,F}\rangle = d^{\dag}_{\mathbf{k}\sigma} |\Phi^{(1)}_{A,F^{\prime}}\rangle=|\Psi_\beta \rangle \otimes |\chi_f(\mathbf{q}\lambda)\rangle \otimes |n_{\mathbf{k}\sigma}^{(d)} \rangle$ with $n_{\mathbf{k}\sigma}^{(d)} = 1\, (0)$ when $\widebar{n}_{\mathbf{k}\sigma}^{(d)} = 0\, (1)$. It should be noted that the time dependent phase factor of $d_{\mathbf{k}\sigma} (t_f)$, $e^{-i\varepsilon_{\mathbf{k}\sigma}^{(d)} t_f/\hbar}$, is irrelevant to the observation value $\langle J_{\mathbf{k}\sigma}\rangle$. Therefore, from Eqs. (\ref{eqnA2.10}) and (\ref{eqnA2.25}), we have the following relation  
\begin{equation}
\widebar{\Gamma}_A^{(1)} \simeq \langle J_{\mathbf{k}\sigma}\rangle = \langle d^\dag_{\mathbf{k}\sigma} d_{\mathbf{k}\sigma} \rangle .  
 \label{eqnA2.26}  
\end{equation} 
This is an interesting result that the photoemission probability $\widebar{\Gamma}_A^{(1)}$ we have introduced for the ARPES measurement is equivalent approximately to the observation value of the photoelectron operator $J_{\mathbf{k}\sigma} = d^\dag_{\mathbf{k}\sigma} d_{\mathbf{k}\sigma}$. It should be noted that the observation value of $J_{\mathbf{k}\sigma}$ has a similar formula to a photoelectron current operator introduced previously in the reference [\onlinecite{DevereauxPRB2023}].

\section{$S$-matrix perturbation theory for post-experiment $\text{c}$ARPES}  \label{SecA3}

The photoemission probability of the ARPES measurement can provide the single-particle spectral function of the target electrons. This is clearly shown in Eq. (\ref{eqnA2.23}). In principle, this stems from the fact that the ARPES detect the photoemission probability of single-photoelectric processes, in each of which there is one target electron annihilated. Therefore, the photoemission probability of the ARPES measurement involves the single-particle physics of the target electrons. A further extension of the measurement principle of the ARPES can be given as follows. When two photoelectric processes are detected in coincidence, the coincidence detection probability will provide the two-body correlations of the target electrons since there are two electrons are annihilated in the two photoelectric processes. This is the basic idea for the proposal of the coincidence angle-resolved photoemission spectroscopy (cARPES) we have provided previously \citep{SuZhang2020}. 

Let us consider the {\it post-experiment} cARPES we have proposed in the main text, which has a same combined system to the ARPES. Suppose the initial photon states from the photon pulses are same to that defined for the ARPES. Two single-photoelectron detectors $D_1$ and $D_2$ detect the emitted photoelectrons from each photon pulse with fixed momenta and spins $(\mathbf{k}_1\sigma_1)$ and $(\mathbf{k}_2\sigma_2)$, respectively. The {\it post-experiment} cARPES can detect the coincidence detection probability of two photoelectrons which come from two photoelectric processes excited by each photon pulse. Since the coincidence detection probability of these {\it pulse}-resolved two photoelectrons defines the coincidence probability of the relevant {\it pulse}-resolved two photoelectric processes, the {\it post-experiment} cARPES can detect the coincidence probability of {\it pulse}-resolved two photoelectric processes, which involves the two-body correlations of the target electrons.  

The coincidence probability of {\it pulse}-resolved two photoelectric processes for the {\it post-experiment} cARPES measurement can be defined by 
\begin{equation}
\widebar{\Gamma}_A^{(2)} = \text{Tr} [\rho_{A,I}^{(2)}(t_f) 1_A^{(2)}] = \text{Tr} [\rho_A S_A^{(2)}(t_i,t_f) 1_A^{(2)} S_A^{(2)}(t_f, t_i)] . \label{eqnA3.1}
\end{equation}     
Here $\rho_{A,I}^{(2)}$ is the second-order part of the density matrix $\rho_{A,I}(t_f)$, $\rho_A$ is defined by Eq. (\ref{eqnA2.4}) with the initial states of the combined system same to that of the ARPES measurement, i.e., $| \Phi^{(2)}_{A,I} \rangle = | \Phi^{(1)}_{A,I} \rangle$. $S_A^{(2)}(t_f,t_i)$ and $S_A^{(2)} (t_i,t_f)$ are the second-order perturbation expansions of the $S_A$-matrices and defined as 
\begin{eqnarray}
&& S_A^{(2)}(t_f, t_i) = \frac{1}{2}(-\frac{i}{\hbar})^2 \iint_{t_i}^{t_f} d t_2 d t_1 T_t [V_{A,I}(t_2) V_{A,I}(t_1)] , \label{eqnA3.2} \\
&& S_A^{(2)}(t_i, t_f) = \frac{1}{2}(-\frac{i}{\hbar})^2 \iint_{t_f}^{t_i} d t_1 d t_2 \widetilde{T}_t [V_{A,I}^\dag (t_1) V_{A,I}^\dag (t_2)] . \label{eqnA3.3}
\end{eqnarray}
In Eq. (\ref{eqnA3.1}), $1_A^{(2)}$ is a projection operator for the final states of the {\it post-experiment} cARPES measurement and defined as 
\begin{equation}
1_A^{(2)} = \sum_{F} | \Phi^{(2)}_{A,F} \rangle \langle \Phi^{(2)}_{A,F} |, \,\, |\Phi^{(2)}_{A,F}\rangle = |\Psi_\beta \rangle \otimes |\chi_f(\mathbf{q}\lambda)\rangle \otimes |n_{\mathbf{k}_1\sigma_1}^{(d)} n_{\mathbf{k}_2\sigma_2}^{(d)}\rangle , \label{eqnA3.4}
\end{equation}
where $\sum_{F}=\sum_{\beta\mathbf{q\lambda}\chi_f n^{(d)}}$, $n_{\mathbf{k}_1\sigma_1}^{(d)}$ and $n_{\mathbf{k}_2\sigma_2}^{(d)}$ are the numbers of the photoelectrons arrived at two single-photoelectron detectors $D_1$ and $D_2$, respectively. 
$\widebar{\Gamma}_A^{(2)}$ can be expressed into another form as  
\begin{equation}
\widebar{\Gamma}_A^{(2)} = \sum_{I} P_{A,I} \langle \Phi^{(2)}_{A,I} (t_f) | 1_A^{(2)} | \Phi^{(2)}_{A,I} (t_f) \rangle , \, \, |\Phi^{(2)}_{A,I} (t_f)\rangle = S_A^{(2)} (t_f, t_i) | \Phi^{(2)}_{A,I} \rangle , \label{eqnA3.5} 
\end{equation}
where $P_{A,I}$ is defined by Eq. (\ref{eqnA2.5-1}). 
Therefore, the coincidence probability of the {\it post-experiment} cARPES measurement can be shown to follow
\begin{equation}
\widebar{\Gamma}_A^{(2)}=\frac{1}{Z}\sum_{I F} e^{-\beta E_\alpha} P_A(\mathbf{q},\lambda) \big| \langle \Phi^{(2)}_{A,F} | S_A^{(2)} (t_f, t_i) | \Phi^{(2)}_{A,I} \rangle \big|^2 , \label{eqnA3.6} 
\end{equation}
where $\sum_{IF}=\sum_{\alpha\beta\mathbf{q\lambda}\chi_i\chi_f n^{(d)}}$. Eq. (\ref{eqnA3.6}) is one main result of the $S$-matrix perturbation theory for the {\it post-experiment} cARPES.  

Let us first consider two photoelectric processes which are caused by one photon pulse with one initial state $| \Phi^{(2)}_{A,I} \rangle$ and one final state $| \Phi^{(2)}_{A,F} \rangle$. The coincidence probability of the two photoelectric processes can be defined by 
\begin{equation}
\widebar{\Gamma}_{A,IF}^{(2)}= \big| \langle \Phi^{(2)}_{A,F} | S_A^{(2)} (t_f, t_i) | \Phi^{(2)}_{A,I} \rangle \big|^2 ,  \label{eqnA3.7} 
\end{equation}
which follows, with the simplified denotations $| \Phi^{(2)}_{A,I} \rangle = |\Psi_\alpha; \chi_{i}(\mathbf{q}\lambda); 0^{(d)}\rangle$ and $| \Phi^{(2)}_{A,F} \rangle = |\Psi_\beta; \chi_{f}(\mathbf{q}\lambda); n_{\mathbf{k}_1\sigma_1}^{(d)} n_{\mathbf{k}_2\sigma_2}^{(d)}\rangle $, 
\begin{eqnarray}
\hspace{-2em}\widebar{\Gamma}_{A,IF}^{(2)} 
&=& \Big| \frac{1}{2}(-\frac{i}{\hbar})^2 \iint_{t_i}^{t_f} d t_2 d t_1 \langle n_{\mathbf{k}_1\sigma_1}^{(d)} n_{\mathbf{k}_2\sigma_2}^{(d)}; \chi_f(\mathbf{q}\lambda); \Psi_\beta| T_t [V_{A,I}(t_2) V_{A,I}(t_1)] |\Psi_\alpha; \chi_{i}(\mathbf{q}\lambda); 0^{(d)}\rangle \Big|^2 \notag \\
&=& \Big| (-\frac{i}{\hbar})^2 \iint_{t_i}^{t_f} d t_2 d t_1 \sum_{\mathbf{k}_1^\prime\sigma_1^\prime \mathbf{k}_2^\prime\sigma_2^\prime} g_{A,1} g_{A,2} \langle \Psi_\beta| T_t c_{\mathbf{k}_2^\prime-\mathbf{q} \sigma_2}(t_2) c_{\mathbf{k}_1^\prime-\mathbf{q} \sigma_1}(t_1) |\Psi_\alpha\rangle  \langle\chi_f(\mathbf{q}\lambda)| a^2_{\mathbf{q}\lambda} |\chi_{i}(\mathbf{q}\lambda) \rangle  \Big. \notag \\
&& \Big. \times \langle n_{\mathbf{k}_1\sigma_1}^{(d)}| d^\dag_{\mathbf{k}_1^\prime\sigma_1^\prime}  |0^{(d)}\rangle
\langle n_{\mathbf{k}_2\sigma_2}^{(d)}| d^\dag_{\mathbf{k}_2^\prime\sigma_2^\prime}  |0^{(d)}\rangle e^{i [\varepsilon_{\mathbf{k}_1^\prime}^{(d)}/\hbar -\omega_{\mathbf{q}}] t_1 + i [\varepsilon_{\mathbf{k}_2^\prime}^{(d)}/\hbar -\omega_{\mathbf{q}}] t_2} \Big|^2 , \label{eqnA3.8}
\end{eqnarray}
where we have taken account of two contributions which are equal under the transformations $(\mathbf{k}_1 \leftrightarrow \mathbf{k}_2, \sigma_1 \leftrightarrow \sigma_2, t_1 \leftrightarrow t_2)$. Here $g_{A,1}= g_{A}(\mathbf{k}_1^\prime-\mathbf{q};\mathbf{q},\lambda_1)$ and $g_{A,2}= g_{A}(\mathbf{k}_2^\prime-\mathbf{q};\mathbf{q},\lambda_2)$. With a similar discussion for $\widebar{\Gamma}_{A,IF}^{(1)}$, we can express $\widebar{\Gamma}_{A,IF}^{(2)}$ into the following simplified form as 
\begin{equation}
\widebar{\Gamma}_{A,IF}^{(2)} = \big| \langle \Phi^{(2)}_{A,F} | \Phi^{(2)}_{\mathbf{k}_1\sigma_1 \mathbf{k}_2\sigma_2} (t_f) \rangle \big|^2 \delta_{\mathbf{k}_1 \mathbf{k}_1^\prime}\delta_{\sigma_1\sigma_1^\prime} \delta_{\mathbf{k}_2 \mathbf{k}_2^\prime}\delta_{\sigma_2\sigma_2^\prime}, \label{eqnA3.9}
\end{equation}
where 
$|\Phi^{(2)}_{\mathbf{k}_1\sigma_1 \mathbf{k}_2\sigma_2} (t_f) \rangle$ is defined by 
\begin{equation}
 |\Phi^{(2)}_{\mathbf{k}_1\sigma_1 \mathbf{k}_2\sigma_2} (t_f) \rangle =  (-\frac{i}{\hbar})^2 \iint_{t_i}^{t_f} d t_2 d t_1  g_{A,1} g_{A,2} T_t [d^\dag_{\mathbf{k}_2\sigma_2}(t_2) c_{\mathbf{k}_2-\mathbf{q} \sigma_2}(t_2) a_{\mathbf{q} \lambda}(t_2) d^\dag_{\mathbf{k}_1\sigma_1}(t_1) c_{\mathbf{k}_1-\mathbf{q} \sigma_1}(t_1) a_{\mathbf{q} \lambda}(t_1) ] |\Phi^{(2)}_{A,I} \rangle . \label{eqnA3.10}
\end{equation}

With a same trick to introduce a photoelectron-state factor for the ARPES measurement, $\widebar{\Gamma}_{A,IF}^{(2)}$ can be expressed into the form as 
\begin{equation}
\widebar{\Gamma}_{A,IF}^{(2)}  = \Gamma_{A,\alpha\beta}^{(2)} \cdot I_{A,\chi}^{(2)} \cdot I_{A,d}^{(2)}, \label{eqnA3.11}  
\end{equation} 
where $\Gamma_{A,\alpha\beta}^{(2)}$ is a target-electron form factor, $I_{A,\chi}^{(2)}$ is a photon-state factor  and $I_{A,d}^{(2)}$ is a photoelectron-state factor. The target-electron form factor $\Gamma_{A,\alpha\beta}^{(2)}$ for the {\it post-experiment} cARPES measurement follows
\begin{eqnarray}
\Gamma_{A,\alpha\beta}^{(2)} &=& \Big| (-\frac{i}{\hbar})^2 \iint_{t_i}^{t_f} d t_2 d t_1 \, g_{A,1} g_{A,2} \langle \Psi_\beta| T_t c_{\mathbf{k}_2-\mathbf{q} \sigma_2}(t_2) c_{\mathbf{k}_1-\mathbf{q} \sigma_1}(t_1) |\Psi_\alpha\rangle  e^{i [\varepsilon_{\mathbf{k}_1}^{(d)}/\hbar -\omega_{\mathbf{q}}] t_1 + i [\varepsilon_{\mathbf{k}_2}^{(d)}/\hbar -\omega_{\mathbf{q}}] t_2} \Big|^2 \notag \\
&=& \frac{|g_{A,1}g_{A,2}|^2}{\hbar^4} \Big| \iint_{t_i}^{t_f} d t_2 d t_1 \, \Phi^{(2)}_{A,\alpha\beta}(\mathbf{k}_1-\mathbf{q} \sigma_1 t_1; \mathbf{k}_2-\mathbf{q} \sigma_2 t_2)  e^{i [\varepsilon_{\mathbf{k}_1}^{(d)}/\hbar -\omega_{\mathbf{q}}] t_1 + i [\varepsilon_{\mathbf{k}_2}^{(d)}/\hbar -\omega_{\mathbf{q}}] t_2} \Big|^2  , \label{eqnA3.12}
\end{eqnarray}
where we have introduced a two-body Bethe-Salpeter wave function in particle-particle channel \citep{SalpeterBethe1951, GellmanLowBS1951}, 
\begin{equation}
\Phi^{(2)}_{A,\alpha\beta}(\mathbf{k}_1 \sigma_1 t_1; \mathbf{k}_2 \sigma_2 t_2) = \langle \Psi_{\beta} | T_t c_{\mathbf{k}_2\sigma_2}(t_2) c_{\mathbf{k}_1\sigma_1} (t_1) | \Psi_{\alpha }\rangle . \label{eqnA3.13}
\end{equation}
Defining the center-of-mass time $t_c = (t_1+t_2)/2$ and the relative time $t_r = t_2 - t_1$, we can introduce another expression for the two-body Bethe-Salpeter wave function, $\Phi^{(2)}_{A,\alpha\beta}(\mathbf{k}_1 \sigma_1, \mathbf{k}_2 \sigma_2; t_c, t_r)=\Phi^{(2)}_{A,\alpha\beta}(\mathbf{k}_1 \sigma_1 t_1; \mathbf{k}_2 \sigma_2 t_2)$. The Fourier transformation of $\Phi^{(2)}_{A,\alpha\beta}(\mathbf{k}_1 \sigma_1, \mathbf{k}_2 \sigma_2; t_c, t_r)$ can be defined as 
\begin{equation}
\Phi^{(2)}_{A,\alpha\beta}(\mathbf{k}_1 \sigma_1, \mathbf{k}_2\sigma_2; \Omega, \omega) =  \iint_{-\infty}^{+\infty} d t_c d t_r \Phi^{(2)}_{A,\alpha\beta}(\mathbf{k}_1 \sigma_1, \mathbf{k}_2\sigma_2; t_c, t_r)  e^{i \Omega t_c + i \omega t_r} .  \label{eqnA3.14}
\end{equation}
Thus, the target-electron form factor $\Gamma_{A,\alpha\beta}^{(2)}$ can be shown to follow
\begin{equation}
\Gamma_{A,\alpha\beta}^{(2)} = \frac{|g_{A,1}g_{A,2}|^2}{\hbar^4}  \big| \Phi_{A,\alpha\beta}^{(2)} (\mathbf{k}_{A,1} \sigma_1,\mathbf{k}_{A,2} \sigma_2;\Omega_A,\omega_A) \big|^2 ,   \label{eqnA3.15}
\end{equation}
where $\mathbf{k}_{A,1} = \mathbf{k}_1-\mathbf{q}$, $\mathbf{k}_{A,2} = \mathbf{k}_2-\mathbf{q}$, $g_{A,1}= g_A(\mathbf{k}_{A,1};\mathbf{q},\lambda)$ and $g_{A,2} = g_A(\mathbf{k}_{A,2};\mathbf{q},\lambda)$. Here we have set $t_i\rightarrow -\infty$ and $t_f \rightarrow +\infty$ in the last step of the derivation. In Eq. (\ref{eqnA3.15}), the center-of-mass frequency $\Omega_A$ and the relative frequency $\omega_A$ are defined by $\Omega_A = (E_{A,1}+E_{A,2})/\hbar, \ \omega_A = (E_{A,2}-E_{A,1})/2\hbar$, where the two transferred energies in the two photoelectric processes are defined by $E_{A,1} = \varepsilon_{\mathbf{k}_1}^{(d)} + \Phi - \hbar \omega_{\mathbf{q}}$ and $E_{A,2} = \varepsilon_{\mathbf{k}_2}^{(d)} + \Phi - \hbar \omega_{\mathbf{q}}$. 
The photon-state factor $I_{A,\chi}^{(2)}$ and the photoelectron-state factor $I_{A,d}^{(2)}$ in Eq. (\ref{eqnA3.11}) for $\widebar{\Gamma}_{A,IF}^{(2)}$ are defined by
\begin{eqnarray}
I_{A,\chi}^{(2)} &=& \big| \langle\chi_f(\mathbf{q}\lambda)| a^2_{\mathbf{q}\lambda} |\chi_{i}(\mathbf{q}\lambda) \rangle \big|^2 , \label{eqnA3.16} \\
I_{A,d}^{(2)} &=& I_{A,d_1}^{(1)} \times I_{A,d_2}^{(1)} , \label{eqnA3.17}
\end{eqnarray}
where the two photoelectron-state factors $I_{A,d_1}^{(1)}$ and $I_{A,d_2}^{(1)}$ are defined by   
\begin{equation}
I_{A,d_1}^{(1)} = \big| \langle n_{\mathbf{k}_1\sigma_1}^{(d)}| d^\dag_{\mathbf{k}_1\sigma_1}  |0^{(d)}\rangle \big|^2, \,\, I_{A,d_2}^{(1)} = \big| \langle n_{\mathbf{k}_2\sigma_2}^{(d)}| d^\dag_{\mathbf{k}_2\sigma_2}  |0^{(d)}\rangle \big|^2 . \label{eqnA3.18}
\end{equation}
Since $I_{A,d_1}^{(1)} = 0 \, (1)$ when $n_{\mathbf{k}_1\sigma_1}^{(d)} = 0 \, (1)$ and $I_{A,d_2}^{(1)} = 0 \, (1)$ when $n_{\mathbf{k}_2\sigma_2}^{(d)} = 0 \, (1)$, $I_{A,d}^{(2)}$ records the coincidence counting of the {\it pulse}-resolved photoelectrons arrived at two single-photoelectron detectors $D_1$ and $D_2$. 

The statistical average of the coincidence probability of two photoelectric processes from every one of the sequential photon pulses can be shown from Eq. (\ref{eqnA3.6}) to follow 
\begin{equation}
\widebar{\Gamma}_A^{(2)} = \frac{1}{Z}\sum_{I F} e^{-\beta E_\alpha} P_{A}(\mathbf{q},\lambda) \cdot \Gamma_{A,\alpha\beta}^{(2)} \cdot I_{A,\chi}^{(2)} \cdot I_{A,d}^{(2)} ,  \label{eqnA3.19}
\end{equation}
where $\sum_{I F} = \sum_{\alpha\beta} \sum_{\mathbf{q}\lambda \chi_i \chi_f } \sum_{n_1^{(d)} n_2^{(d)}}$. It should be remarked that $\widebar{\Gamma}_{A,IF} ^{(2)}$ has a same structure to $\widebar{\Gamma}^{(2)}$ in Eq. (1) of the main text, i.e., $\widebar{\Gamma}_{A,IF}^{(2)}  = \Gamma_{A,\alpha\beta}^{(2)} \cdot I_{A,\chi}^{(2)} \cdot I_{A,d}^{(2)}$, and $I_{A,d}^{(2)} = I_{A,d_1}^{(1)} \times  I_{A,d_2}^{(1)}$ follows in both $\widebar{\Gamma}_{A,IF}^{(2)}$ and $\widebar{\Gamma}_A^{(2)}$. This shows that the coincidence probability of {\it pulse}-resolved two photoelectric processes can be obtained by $I_{A,d}^{(2)}$ which records the coincidence counting of the {\it pulse}-resolved photoelectrons arrived at two single-photoelectron detectors renormalized by the target-electron form factor and the photon-state factor. Therefore, a {\it post-experiment} cARPES can be designed following the {\it post-experiment} coincidence counting method we have presented in the main text. It is noted that when $I_{A,d_1}^{(1)} = 1$ and $I_{A,d_2}^{(1)} = 1$, $\widebar{\Gamma}_A^{(2)}$ recovers the previous results of the {\it instantaneous} cARPES we have proposed previously \citep{SuZhang2020}. 

Now let us give a simple discussion on the two-body Bethe-Salpeter wave function $\Phi^{(2)}_{A,\alpha\beta}$. From Eq. (\ref{eqnA3.13}), it is clear that the two-body Bethe-Salpeter wave function for the cARPES describes the dynamical physics of the target electrons when two electrons are annihilated in time ordering, thus it describes the dynamical two-body correlations of the target electrons in particle-particle channel. The frequency Bethe-Salpeter wave function has a general form \citep{SuZhang2020}  
\begin{equation}
\phi^{(2)}_{A,\alpha\beta}\left(\mathbf{k}_1\sigma_1, \mathbf{k}_2\sigma_2; \Omega, \omega \right)  = 2\pi \delta \left[\Omega + \left( E_{\beta} - E_{\alpha}\right)/\hbar \right] \phi^{(2)}_{A,\alpha\beta}\left(\mathbf{k}_1\sigma_1, \mathbf{k}_2\sigma_2; \omega \right) , \label{eqnA3.20} 
\end{equation} 
where $\phi^{(2)}_{A,\alpha\beta}\left(\mathbf{k}_1\sigma_1, \mathbf{k}_2\sigma_2; \omega \right)$ follows
\begin{equation}
\phi^{(2)}_{A,\alpha\beta}\left(\mathbf{k}_1\sigma_1, \mathbf{k}_2\sigma_2; \omega \right) =\sum_{\gamma} \left[ \frac{ i \langle\Psi_{\beta} \vert c_{\mathbf{k}_2 \sigma_2} \vert \Psi_{\gamma} \rangle \langle \Psi_{\gamma} \vert  c_{\mathbf{k}_1 \sigma_1} \vert \Psi_{\alpha} \rangle} {\omega + i\delta^+ + (E_{\alpha} + E_{\beta} - 2 E_{\gamma} )/2\hbar} + \frac{ i \langle\Psi_{\beta} \vert c_{\mathbf{k}_1 \sigma_1} \vert \Psi_{\gamma} \rangle \langle \Psi_{\gamma} \vert  c_{\mathbf{k}_2 \sigma_2} \vert \Psi_{\alpha} \rangle} {\omega - i\delta^+ - (E_{\alpha} + E_{\beta} - 2 E_{\gamma} )/2\hbar} \right] . \label{eqnA3.21}
\end{equation} 
The frequency Bethe-Salpeter wave function involves the following physics \citep{SuZhang2020}: (1) The center-of-mass dynamical physics of two target electrons described by the $\delta$-function, $\delta \left[\Omega + \left( E_{\beta} - E_{\alpha}\right)/\hbar \right]$, which shows the energy transfer conservation  in the center-of-mass channel; (2) The inner-pair dynamical physics described by 
$\phi^{(2)}_{\alpha\beta}\left(\mathbf{k}_1\sigma_1, \mathbf{k}_2\sigma_2; \omega \right)$, which shows the propagatorlike resonance structures, peaked at $\omega = \pm (E_{\alpha} + E_{\beta} - 2 E_{\gamma} )/2\hbar$ with the weights defined by $\langle\Psi_{\beta} \vert c_{\mathbf{k}_2 \sigma_2} \vert \Psi_{\gamma} \rangle \langle \Psi_{\gamma} \vert  c_{\mathbf{k}_1 \sigma_1} \vert \Psi_{\alpha} \rangle$ and $\langle\Psi_{\beta} \vert c_{\mathbf{k}_1 \sigma_1} \vert \Psi_{\gamma} \rangle \langle \Psi_{\gamma} \vert  c_{\mathbf{k}_2 \sigma_2} \vert \Psi_{\alpha} \rangle$. The spectral function of the two-body Bethe-Salpeter wave function $\phi^{(2)}_{A,\alpha\beta}\left(\mathbf{k}_1\sigma_1, \mathbf{k}_2\sigma_2; \Omega, \omega \right)$ shows us that the cARPES can provide the dynamical two-body correlations of the target electrons, which include both the center-of-mass and the inner-pair relative dynamical physics with both energy and momentum resolved.  

Following the discussion on the approximate equivalence of the photoemission probability of the ARPES measurement $\widebar{\Gamma}_{A}^{(1)}$ and the observation value of the photoelectron operator $J_{\mathbf{k}\sigma} = d^\dag_{\mathbf{k}\sigma} d_{\mathbf{k}\sigma}$ as shown in Eq. (\ref{eqnA2.26}), we now make a similar discussion on the relation between the coincidence probability $\widebar{\Gamma}_{A}^{(2)}$ and a pair-photoelectron operator for the cARPES measurement $J_{\mathbf{k}_1\sigma_1 \mathbf{k}_2\sigma_2}=d^\dag_{\mathbf{k}_2\sigma_2} d_{\mathbf{k}_2\sigma_2} d^\dag_{\mathbf{k}_1\sigma_1} d_{\mathbf{k}_1\sigma_1}$. Here the operators $d^\dag_{\mathbf{k}_1\sigma_1} \, (d_{\mathbf{k}_1\sigma_1})$ and $d^\dag_{\mathbf{k}_2\sigma_2} \, (d_{\mathbf{k}_2\sigma_2})$ are defined for the two single-photoelectron detectors $D_1$ and $D_2$, respectively. The statistical ensemble average of $J_{\mathbf{k}_1\sigma_1 \mathbf{k}_2\sigma_2}$ at the observation time $t_f$ can be defined following Eq. (\ref{eqnA1.20}), $\langle J_{\mathbf{k}_1\sigma_1 \mathbf{k}_2\sigma_2}\rangle = \text{Tr}[\rho_A S_A(t_i, t_f) J_{I,\mathbf{k}_1\sigma_1 \mathbf{k}_2\sigma_2} (t_f) S_A(t_f, t_i)]$. In the cARPES coincidence detection, it is the the second-order perturbation  expansions of the $S$-matrices that have main contributions to the observation value $\langle J_{\mathbf{k}_1\sigma_1 \mathbf{k}_2\sigma_2}\rangle$. Therefore, $\langle J_{\mathbf{k}_1\sigma_1 \mathbf{k}_2\sigma_2}\rangle$ can be calculated approximately by
\begin{eqnarray}
\langle J_{\mathbf{k}_1\sigma_1 \mathbf{k}_2\sigma_2}\rangle  &\simeq& \text{Tr}[\rho_A S_A^{(2)}(t_i, t_f) d^{\dag}_{\mathbf{k}_1\sigma_1}(t_f) d^{\dag}_{\mathbf{k}_2\sigma_2} (t_f) d_{\mathbf{k}_2\sigma_2}(t_f) d_{\mathbf{k}_1\sigma_1}(t_f) S_A^{(2)}(t_f, t_i)]  \notag \\
& = & \text{Tr}[\rho_A S_A^{(2)}(t_i, t_f) d^{\dag}_{\mathbf{k}_1\sigma_1}(t_f) d^{\dag}_{\mathbf{k}_2\sigma_2} (t_f) \, 1_A^{\prime (2)} \, d_{\mathbf{k}_2\sigma_2}(t_f) d_{\mathbf{k}_1\sigma_1}(t_f) S_A^{(2)}(t_f, t_i)]  \notag \\
&=& \sum_{I F^\prime} P_{A,I}  \langle \Phi^{(2)}_{A,I} | S_A^{(2)} (t_i, t_f) d^{\dag}_{\mathbf{k}_1\sigma_1}(t_f) d^{\dag}_{\mathbf{k}_2\sigma_2} (t_f) | \Phi^{(2)}_{A,F^\prime} \rangle \langle \Phi^{(2)}_{A,F^{\prime}} | d_{\mathbf{k}_2\sigma_2}(t_f) d_{\mathbf{k}_1\sigma_1}(t_f) S_A^{(2)} (t_f, t_i) | \Phi^{(2)}_{A,I} \rangle  \notag \\
&=& \sum_{I F^\prime} P_{A,I} \big| \langle \Phi^{(2)}_{A,F^{\prime}} | d_{\mathbf{k}_2\sigma_2} d_{\mathbf{k}_1\sigma_1} S_A^{(2)} (t_f, t_i) | \Phi^{(2)}_{A,I} \rangle \big|^2 \notag \\
&=& \frac{1}{Z}\sum_{I F^\prime} e^{-\beta E_\alpha} P_A(\mathbf{q},\lambda) \big| \langle \Phi^{(2)}_{A,F} | S_A^{(2)} (t_f, t_i) | \Phi^{(2)}_{A,I} \rangle \big|^2 . \label{eqnA3.22} 
\end{eqnarray}  
Here $1_A^{\prime (2)}=\sum_{F^{\prime}} |\Phi^{(2)}_{A,F^{\prime}}\rangle \langle \Phi^{(2)}_{A,F^{\prime}}|$ with $\sum_{F^\prime}=\sum_{\beta\mathbf{q}\lambda\chi_f\widebar{n}^{(d)}}$ and $|\Phi^{(2)}_{A,F^{\prime}}\rangle = |\Psi_\beta \rangle \otimes |\chi_f(\mathbf{q}\lambda)\rangle \otimes |\widebar{n}_{\mathbf{k}_1\sigma_1}^{(d)} \widebar{n}_{\mathbf{k}_2\sigma_2}^{(d)}\rangle$, and $|\Phi^{(2)}_{A,F}\rangle = d^{\dag}_{\mathbf{k}_1\sigma_1} d^{\dag}_{\mathbf{k}_2\sigma_2} |\Phi^{(2)}_{A,F^{\prime}}\rangle=|\Psi_\beta \rangle \otimes |\chi_f(\mathbf{q}\lambda)\rangle \otimes |n_{\mathbf{k}_1\sigma_1}^{(d)} n_{\mathbf{k}_2\sigma_2}^{(d)}\rangle$ with $n_{\mathbf{k}\sigma}^{(d)} = 1\, (0)$ when $\widebar{n}_{\mathbf{k}\sigma}^{(d)} = 0\, (1)$. From Eqs. (\ref{eqnA3.6}) and (\ref{eqnA3.22}), we can show that 
\begin{equation}
\widebar{\Gamma}_A^{(2)} \simeq \langle J_{\mathbf{k}_1\sigma_1 \mathbf{k}_2\sigma_2}\rangle = \langle d^\dag_{\mathbf{k}_2\sigma_2} d_{\mathbf{k}_2\sigma_2} d^\dag_{\mathbf{k}_1\sigma_1} d_{\mathbf{k}_1\sigma_1} \rangle . 
 \label{eqnA3.23}  
\end{equation} 
This is a very interesting result that the coincidence probability $\widebar{\Gamma}_A^{(2)}$ we have introduced for the cARPES measurement is equivalent approximately to the observation value of the pair-photoelectron operator $J_{\mathbf{k}_1\sigma_1 \mathbf{k}_2\sigma_2}=d^\dag_{\mathbf{k}_2\sigma_2} d_{\mathbf{k}_2\sigma_2} d^\dag_{\mathbf{k}_1\sigma_1} d_{\mathbf{k}_1\sigma_1}$.  Here the pair-photoelectron operator $J_{\mathbf{k}_1\sigma_1 \mathbf{k}_2\sigma_2}$ is closely related to a pair-photoelectron current operator introduced in the reference [\onlinecite{DevereauxPRB2023}] for the cARPES measurement. It should also be noted that the time dependent phase factors of $d_{\mathbf{k}_1\sigma_1} (t_f)$ and $d_{\mathbf{k}_2\sigma_2} (t_f)$ are irrelevant to the observation value $\langle J_{\mathbf{k}_1\sigma_1 \mathbf{k}_2\sigma_2}\rangle$, which implies that the coincidence detection of the emitted photoelectrons from the {\it pulse}-resolved two photoelectric processes is not necessary at {\it simultaneous} time. 

Now let us give another formulation for the coincidence probability of the cARPES measurement. We introduce a two-body non-equilibrium Green's function as 
\begin{equation}
G_c(\mathbf{k}_1\sigma_1, \mathbf{k}_2\sigma_2 ; t_1, t_2 ; t_2^\prime, t_1^\prime) = (i)^2 \langle T_c c^\dag_{\mathbf{k}_1\sigma_1} (t_1^\prime) c^\dag_{\mathbf{k}_2\sigma_2} (t_2^\prime) c_{\mathbf{k}_2\sigma_2} (t_2) c_{\mathbf{k}_1\sigma_1} (t_1) \rangle , \label{eqnA3.24}
\end{equation}
where $\langle A \rangle = \frac{1}{Z} \text{Tr}(e^{-\beta H_s} A)$ with $H_s$ being the target-electron Hamiltonian. Here $T_c$ is a contour-time ordering operator defined on the time contour $C=C_{+} \cup C_{-}$, where $t\in C_{+}$ evolves as $t_i\rightarrow t_f$ and $t^\prime \in C_{-}$ evolves as $t_f\rightarrow t_i$. The definition of $T_c$ is given by \citep{SuZhang2020,Rammer,SuRaman2016}
\begin{equation}
T_c [A(t_1) B(t_2)] = \left\{
\begin{array} {l l l}
A(t_1) B(t_2) , &  \text{if} & t_1 >_c t_2 , \\
\pm B(t_2) A(t_1) , &  \text{if} & t_1 <_c t_2 ,
\end{array}
\right. \label{eqnA3.25}
\end{equation}
where $>_c$ and $<_c$ are defined according to the positions of the time arguments in the time contour $C$, and $\pm$ are defined for the bosonic or fermionic operators, respectively. From the definition of $\widebar{\Gamma}_A^{(2)}$ in Eq. (\ref{eqnA3.5}), we have 
\begin{equation}
\widebar{\Gamma}_A^{(2)} = \sum_{I} P_{A,I} \langle \Phi^{(2)}_{A,I} | S_A^{(2)}(t_i, t_f)\, 1_A^{(2)} \, S_A^{(2)}(t_f, t_i)| \Phi^{(2)}_{A,I} \rangle . \label{eqnA3.26} 
\end{equation}
It can be shown that $\widebar{\Gamma}_A^{(2)}$ follows
\begin{equation}
\widebar{\Gamma}_A^{(2)} = \frac{(-i)^2}{\hbar^4} \sum_{I F} P_{A}(\mathbf{q}\lambda) |g_{A,1} g_{A,2}|^2 \int_{[t_1^\prime, t_2^\prime, t_2, t_1]}  G_c(\mathbf{k}_{A,1}\sigma_1, \mathbf{k}_{A,2}\sigma_2 ; t_1, t_2 ; t_2^\prime, t_1^\prime) e^{i [ E_{A,1} (t_1 - t_1^\prime) + E_{A,2} (t_2 - t_2^\prime)] /\hbar} \, I_{A,\chi}^{(2)} \cdot I_{A,d}^{(2)} ,  \label{eqnA3.27} 
\end{equation}
where $\sum_{I F} = \sum_{\mathbf{q}\lambda \chi_i \chi_f } \sum_{n_1^{(d)} n_2^{(d)}}$,  $\mathbf{k}_{A,1} = \mathbf{k}_1-\mathbf{q}$, $\mathbf{k}_{A,2} = \mathbf{k}_2-\mathbf{q}$, $g_{A,1} = g_A(\mathbf{k}_{A,1};\mathbf{q},\lambda)$, $g_{A,2} = g_A(\mathbf{k}_{A,2};\mathbf{q},\lambda)$, $E_{A,1} = \varepsilon_{\mathbf{k}_1}^{(d)} + \Phi - \hbar \omega_{\mathbf{q}}$ and $E_{A,2} = \varepsilon_{\mathbf{k}_2}^{(d)} + \Phi - \hbar \omega_{\mathbf{q}}$. The contour-time integral is defined by 
\begin{equation}
\int_{[t_1^\prime, t_2^\prime, t_2, t_1]}= \iint_{t_f}^{t_i} d t_1^\prime d t_2^\prime \iint_{t_i}^{t_f} d t_2 d t_1 . \label{eqnA3.28}
\end{equation}
Let us introduce the non-equilibrium frequency Green's function as 
\begin{equation}
G_c(\mathbf{k}_1\sigma_1, \mathbf{k}_2\sigma_2 ; \omega_1, \omega_2; \omega_2^\prime, \omega_1^\prime) =  \int_{[t_1^\prime, t_2^\prime, t_2, t_1]} G_c(\mathbf{k}_1\sigma_1, \mathbf{k}_2\sigma_2; t_1, t_2 ; t_2^\prime, t_1^\prime) e^{i (\omega_1 t_1 +\omega_2 t_2 - \omega_2^\prime t_2^\prime - \omega_1^\prime t_1^\prime ) } . \label{eqnA3.29}
\end{equation} 
$\widebar{\Gamma}_A^{(2)}$ can be expressed into the form as  
\begin{equation}
\widebar{\Gamma}_A^{(2)} = \frac{(-i)^2}{\hbar^4} \sum_{I F} P_{A}(\mathbf{q}\lambda) |g_{A,1} g_{A,2}|^2 \, G_c(\mathbf{k}_1\sigma_1, \mathbf{k}_2\sigma_2 ; \omega_{A,1}, \omega_{A,2}; \omega_{A,2}, \omega_{A,1}) \cdot I_{A,\chi}^{(2)} \cdot I_{A,d}^{(2)} , \label{eqnA3.30}  
\end{equation}
where $\omega_{A,1} = E_{A,1}/\hbar$ and $\omega_{A,2} = E_{A,2}/\hbar$. Therefore, $\widebar{\Gamma}_A^{(2)}$ involves a two-body dynamical non-equilibrium Green's function of the target electrons, as has been pointed previously \citep{SuZhang2020,DevereauxPRB2023}.

\section{$S$-matrix perturbation theory for INS} \label{SecA4}

The combined system for the inelastic neutron scattering (INS) measurement involves the target-electron spin system and the neutrons. The Hamiltonian for the INS measurement is defined by
\begin{equation}
H_B = H_{B,0} + V_B , \label{eqnA4.1}
\end{equation}
where $H_{B,0} = H_s + H_n$ with $H_s$ being the Hamiltonian of the target-electron spin system and $H_n$ being the Hamiltonian of the neutrons. $H_n=\sum_{\mathbf{q}\sigma} \mathcal{E}(\mathbf{q}) f^{\dag}_{\mathbf{q} \sigma} f_{\mathbf{q} \sigma} $, where $f^{\dag}_{\mathbf{q} \sigma}$ and $f_{\mathbf{q} \sigma}$ are the neutron creation and annihilation operators with momentum $\mathbf{q}$ and spin $\sigma$. The electron-neutron spin interaction $V_B$ is given by \citep{SucINS2021,Lovesey1984,Squires1996,FelixPrice2013}
\begin{equation}
V_B = \sum_{\mathbf{q}_i \mathbf{q}_f \sigma_i \sigma_f} g_B(\mathbf{q}) f^{\dag}_{\mathbf{q}_f \sigma_f} \boldsymbol{\tau}_{\sigma_f \sigma_i} f_{\mathbf{q}_i \sigma_i} \cdot \mathbf{S}_{\perp}(\mathbf{q})  \label{eqnA4.2}
\end{equation}
with $\mathbf{q} = \mathbf{q}_f - \mathbf{q}_i$. 
Here $\boldsymbol{\tau}$ is the Pauli matrix and $\mathbf{S}_{\perp}(\mathbf{q})$ is a target-electron spin relevant operator. $\mathbf{S}_{\perp}(\mathbf{q})$ is defined as $\mathbf{S}_{\perp}(\mathbf{q}) =  \mathbf{S}(\mathbf{q}) \cdot (1 - \widehat{\mathbf{q}}\widehat{\mathbf{q}})$, where $\mathbf{S}(\mathbf{q}) = \sum_{l} \mathbf{S}_l e^{-i\mathbf{q}\cdot \mathbf{R}_l}$ with $\mathbf{S}_l$ being the target-electron spin operator at position $\mathbf{R}_l$ and $\widehat{\mathbf{q}} = \mathbf{q}/|\mathbf{q}|$.
The electron-neutron scattering $S$-matrix is defined by
\begin{equation}
S_B (t_f, t_i) = T_t \exp [-\frac{i}{\hbar} \int_{t_i}^{t_f} dt V_{B,I}(t)] , \,\, V_{B,I}(t)= e^{\frac{i}{\hbar} H_{B,0} t} V_B e^{-\frac{i}{\hbar} H_{B,0} t} . \label{eqnA4.3} 
\end{equation}

Suppose the initial states of the combined system at the beginning time $t_i$ of every neutron-scattering process of the INS measurement are defined by an ensemble density matrix
\begin{equation}
\rho_B = \sum_{I} P_{B,I}^{(1)} | \Phi_{B,I}^{(1)} \rangle \langle \Phi_{B,I}^{(1)} | , \label{eqnA4.4}
\end{equation}
where the distribution function $P_{B,I}^{(1)}$ and the initial states $| \Phi_{B,I}^{(1)} \rangle$ are defined by
\begin{eqnarray}
&& P_{B,I}^{(1)} = \frac{1}{Z} e^{-\beta E_\alpha} P_{B}^{(1)}(\mathbf{q}_i, \sigma_i), \label{eqnA4.5-1} \\
&& |\Phi^{(1)}_{B,I} \rangle = | \Psi_{\alpha} \rangle \otimes | n_{\mathbf{q}_i\sigma_i} \rangle . \label{eqnA4.5-2}  
\end{eqnarray}
Here $\sum_{I} = \sum_{\alpha \mathbf{q}_i \sigma_i n_i}$, $Z=\text{Tr} (e^{-\beta H_s})$ and $|\Psi_\alpha \rangle$ are the eigenstates of the target-electron spin system with the corresponding eigenvalues $E_\alpha$, $n_{\mathbf{q}_i\sigma_i}=0, 1$ is the incident neutron number, and $P_{B}^{(1)}(\mathbf{q}_i, \sigma_i)$ is the incident neutron distribution function. It is assumed that the neutron momentum and spin degrees of freedom are decoupled and $P_{B}^{(1)}(\mathbf{q}_i, \sigma_i) = P_{B}^{(1)}(\mathbf{q}_i) \cdot \widebar{P}_{B}^{(1)}(\sigma_i)$, where the incident neutron spins are in the thermal mixed states defined by 
\begin{equation}
\sum_{\sigma_i} \widebar{P}_{B}^{(1)} (\sigma_i) \vert \sigma_i \rangle \langle \sigma_i \vert = \frac{1}{2} \left( \vert \uparrow \rangle \langle \uparrow \vert + \vert \downarrow \rangle \langle \downarrow \vert \right) . \label{eqnA4.6}
\end{equation}
The final states of the scattered neutrons which arrive at the single-neutron detector are assumed with fixed momentum but arbitrary spin. Following the above discussions for the ARPES and the cARPES, we can define a projection operator for the final states of the neutron-scattering processes of the INS measurement as  
\begin{equation}
1_B^{(1)} = \sum_F |\Phi^{(1)}_{B,F} \rangle \langle \Phi^{(1)}_{B,F} | , \,\, 
 |\Phi^{(1)}_{B,F} \rangle = | \Psi_{\beta} \rangle \otimes | n_{\mathbf{q}_f\sigma_f} \rangle , \label{eqnA4.7}
\end{equation}
where $\sum_F = \sum_{\beta\sigma_f n_f}$, and $n_{\mathbf{q}_f\sigma_f}=0, 1$ is defined for the scattered neutron final states. 
  
From Eq. (\ref{eqnA1.18}), the density matrix of the combined system for the INS measurement at the observation time $t_f$ in the interaction picture is given by 
\begin{equation}
\rho_{B,I}(t_f) = S_B(t_f, t_i) \rho_B S_B (t_i, t_f) . \label{eqnA4.8}
\end{equation}   
Since the scattering probability of the INS measurement is dominated by the single neutron-scattering processes, it can be defined by 
\begin{equation}
\widebar{\Gamma}_B^{(1)} = \text{Tr} [\rho_{B,I}^{(1)}(t_f) 1_B^{(1)}] = \text{Tr} [\rho_B S_B^{(1)}(t_i,t_f) 1_B^{(1)} S_B^{(1)}(t_f, t_i) ] , \label{eqnA4.9}
\end{equation}     
where $\rho_{B,I}^{(1)}(t_f)$ is the first-order part of the density matrix $\rho_{B,I}(t_f)$, $S_B^{(1)}(t_f,t_i)$ is the first-order perturbation expansion of the $S_B$-matrix and defined as $S_B^{(1)} (t_f,t_i)=-\frac{i}{\hbar} \int_{t_i}^{t_f} d t V_{B,I}(t)$, and $S_B^{(1)} (t_i,t_f) = S_B^{(1)\dag}(t_f,t_i)$. From a similar discussion for the ARPES, $\widebar{\Gamma}_B^{(1)}$ can be expressed into the following form as
\begin{equation}
\widebar{\Gamma}_B^{(1)} = \sum_{I} P_{B,I}^{(1)} \langle \Phi^{(1)}_{B,I} (t_f) | 1_B^{(1)} | \Phi^{(1)}_{B,I}(t_f) \rangle ,  \label{eqnA4.10} 
\end{equation}
where $|\Phi^{(1)}_{B,I} (t_f)\rangle = S_B^{(1)} (t_f, t_i) | \Phi^{(1)}_{B,I} \rangle$. Therefore, the scattering probability of the single neutron-scattering processes for the INS measurement follows 
\begin{equation}
\widebar{\Gamma}_B^{(1)}=\frac{1}{Z}\sum_{I F} e^{-\beta E_\alpha} P_B^{(1)}(\mathbf{q}_i,\sigma_i) \big| \langle \Phi^{(1)}_{B,F} | S_B^{(1)} (t_f, t_i) | \Phi^{(1)}_{B,I} \rangle \big|^2 , \label{eqnA4.11} 
\end{equation}
where $\sum_{IF}=\sum_{\alpha\beta\mathbf{q}_i\sigma_i\sigma_f n_i n_f}$. Eq. (\ref{eqnA4.11}) is one main result of the $S$-matrix perturbation theory for the INS. 
  
Let us consider one single neutron-scattering process with one initial state $| \Phi^{(1)}_{B,I} \rangle$ and one final state $| \Phi^{(1)}_{B,F} \rangle$. The scattering probability of this single neutron-scattering process is defined by 
\begin{equation}
\widebar{\Gamma}_{B,IF}^{(1)}= \big| \langle \Phi^{(1)}_{B,F} | S_B^{(1)} (t_f, t_i) | \Phi^{(1)}_{B,I} \rangle \big|^2 .  \label{eqnA4.12} 
\end{equation} 
$\widebar{\Gamma}_{B,IF}^{(1)}$ can be calculated as following:  
\begin{eqnarray}
\hspace{-2em} \widebar{\Gamma}_{B,IF}^{(1)} &=& \Big| (-\frac{i}{\hbar}) \int_{t_i}^{t_f} dt  \langle n_{\mathbf{q}_f\sigma_f}; \Psi_\beta | V_{B,I}(t) | \Psi_\alpha; n_{\mathbf{q}_i\sigma_i}\rangle  \Big|^2 \notag \\
&=& \Big| (-\frac{i}{\hbar}) \int_{t_i}^{t_f} dt \sum_{\mathbf{q}_i^\prime \mathbf{q}_f^\prime \sigma_i^\prime \sigma_f^\prime} g_B(\mathbf{q}^\prime) \langle \Psi_\beta | \mathbf{S}_{\perp}(\mathbf{q}^\prime, t)  | \Psi_\alpha \rangle \cdot \boldsymbol{\tau}_{\sigma_f^\prime \sigma_i^\prime} \, \langle n_{\mathbf{q}_f\sigma_f} | f^{\dag}_{\mathbf{q}_f^\prime \sigma_f^\prime} (t) | 0 \rangle  \langle 0 | f_{\mathbf{q}_i^\prime \sigma_i^\prime} (t) | n_{\mathbf{q}_i\sigma_i}\rangle  \Big|^2 \notag \\
&=& \frac{|g_B(\mathbf{q})|^2}{\hbar^2} \Big| \langle \Psi_\beta | \mathbf{S}_{\perp}(\mathbf{q})  | \Psi_\alpha \rangle \cdot \boldsymbol{\tau}_{\sigma_f \sigma_i} \, \langle n_{\mathbf{q}_f \sigma_f} | f^{\dag}_{\mathbf{q}_f \sigma_f} | 0 \rangle \langle 0 | f_{\mathbf{q}_i \sigma_i} | n_{\mathbf{q}_i\sigma_i}\rangle \int_{t_i}^{t_f} dt e^{i [E_\beta -E_\alpha + E_B^{(1)}] t/\hbar} \Big|^2 . \label{eqnA4.13}
\end{eqnarray}
In the last step, we have used the trick to introduce the variables $\langle 0 | f_{\mathbf{q}_i\sigma_i} | n_{\mathbf{q}_i\sigma_i} \rangle$ and $\langle n_{\mathbf{q}_f\sigma_f} | f^\dag_{\mathbf{q}_f\sigma_f} | 0 \rangle$ to describe the incident and the scattered neutron states. Here $E_{B}^{(1)}$ is the transferred energy defined by $E_{B}^{(1)} = \mathcal{E}(\mathbf{q}_{f}) - \mathcal{E}(\mathbf{q}_{i})$, where $\mathcal{E}(\mathbf{q}_{i})$ and $\mathcal{E}(\mathbf{q}_{f})$ are the incident and the scattered neutron energies, respectively. With a similar derivation of Eq. (\ref{eqnA2.17}), $\widebar{\Gamma}_{B,IF}^{(1)}$ can be shown to follow  
\begin{equation}
\widebar{\Gamma}_{B,IF}^{(1)} = \Gamma_{B,\alpha\beta}^{(1)} \cdot I_{B,\chi}^{(1)} \cdot I_{B,d}^{(1)} ,  \label{eqn14}
\end{equation}
where $\Gamma_{B,\alpha\beta}^{(1)}$ is a target-electron spin form factor given by 
\begin{equation}
\Gamma_{B,\alpha\beta}^{(1)} = \frac{2\pi |g_B(\mathbf{q})|^2 \Delta t_d}{\hbar}   |\langle \Psi_\beta | \mathbf{S}_{\perp}(\mathbf{q})  | \Psi_\alpha \rangle \cdot \boldsymbol{\tau}_{\sigma_f \sigma_i}|^2 \, \delta(E_\beta -E_\alpha + E_B^{(1)}) , \label{eqnA4.15} 
\end{equation}
$I_{B,\chi}^{(1)}$ is an incident-neutron-state factor and $I_{B,d}^{(1)}$ is a scattered-neutron-state factor, which are defined by 
\begin{eqnarray}
I_{B,\chi}^{(1)} &=& \big| \langle 0 | f_{\mathbf{q}_i\sigma_i} | n_{\mathbf{q}_i\sigma_i} \rangle \big|^2 , \label{eqnA4.16-1} \\ 
I_{B,d}^{(1)} &=& \big| \langle n_{\mathbf{q}_f\sigma_f} | f^\dag_{\mathbf{q}_f\sigma_f} | 0 \rangle \big|^2 . \label{eqnA4.16-2}
\end{eqnarray}
It is noted that $I_{B,\chi}^{(1)} = 0 \, (1)$ when $n_{\mathbf{q}_i\sigma_i} = 0 \, (1)$ and $I_{B,d}^{(1)} = 0 \, (1)$ when $n_{\mathbf{q}_f\sigma_f} = 0 \, (1)$. The scattered-neutron-state factor plays a role to record the number of the neutrons arrived at the single-neutron detector.   

The statistical average of the single neutron-scattering probability for the INS measurement can be calculated from Eq. (\ref{eqnA4.11}), which follows 
\begin{equation}
\widebar{\Gamma}_B^{(1)}=\frac{1}{Z}\sum_{I F} e^{-\beta E_\alpha} P_B^{(1)}(\mathbf{q}_i,\sigma_i) \cdot \Gamma_{B,\alpha\beta}^{(1)} \cdot I_{B,\chi}^{(1)} \cdot I_{B,d}^{(1)} , \label{eqnA4.17} 
\end{equation}  
where $\sum_{IF}=\sum_{\alpha\beta\mathbf{q}_i \sigma_i \sigma_f n_i n_f}$. Let us first consider the sum over the spins $\sigma_i$ and $\sigma_f$ as following: 
\begin{eqnarray}
\widetilde{\Gamma}_{B,IF}^{(1)} &\equiv &  \sum_{\sigma_i \sigma_f} \widebar{P}_{B}^{(1)} (\sigma_i) |\langle \Psi_\beta | \mathbf{S}_{\perp}(\mathbf{q})  | \Psi_\alpha \rangle \cdot \boldsymbol{\tau}_{\sigma_f \sigma_i}|^2  \cdot I_{B,\chi}^{(1)} \cdot I_{B,d}^{(1)} \notag \\
&=& \sum_{ij} \sum_{\sigma_i \sigma_f} \widebar{P}_{B}^{(1)} (\sigma_i) \langle \Psi_\alpha | S^{(i) \dag}_{\perp}(\mathbf{q})  | \Psi_\beta \rangle \langle \Psi_\beta | S^{(j)}_{\perp}(\mathbf{q})  | \Psi_\alpha \rangle \boldsymbol{\tau}^{i}_{\sigma_i \sigma_f}   \boldsymbol{\tau}^{j}_{\sigma_f \sigma_i}   \cdot I_{B,\chi}^{(1)} \cdot I_{B,d}^{(1)} \notag \\
&=& \sum_{ij} \langle \Psi_\alpha | S^{(i) \dag}_{\perp}(\mathbf{q})  | \Psi_\beta \rangle \langle \Psi_\beta | S^{(j)}_{\perp}(\mathbf{q})  | \Psi_\alpha \rangle \sum_{\sigma_i \sigma_f} \frac{1}{2} \langle \sigma_i | \boldsymbol{\tau}^{i} | \sigma_f \rangle  
\langle \sigma_f | \boldsymbol{\tau}^{j} | \sigma_i \rangle \cdot I_{B,\chi}^{(1)} \cdot I_{B,d}^{(1)} \notag \\
&=& \sum_{i} \langle \Psi_\alpha | S^{(i) \dag}_{\perp}(\mathbf{q})  | \Psi_\beta \rangle \langle \Psi_\beta | S^{(i)}_{\perp}(\mathbf{q})  | \Psi_\alpha \rangle \cdot I_{B,\chi}^{(1)} \cdot I_{B,d}^{(1)} . \label{eqnA4.18}
\end{eqnarray} 
Here we have used $\sum_{\sigma_i \sigma_f} \frac{1}{2} \langle \sigma_i | \boldsymbol{\tau}^{i} | \sigma_f \rangle   \langle \sigma_f | \boldsymbol{\tau}^{j} | \sigma_i \rangle = \delta_{i j}$. Another trick in the last step for Eq. (\ref{eqnA4.18}) is based on the fact that $I_{B,\chi}^{(1)}, I_{B,d}^{(1)} = 0 $ or $1$. Therefore, only the terms with $I_{B,\chi}^{(1)}=1$ and $I_{B,d}^{(1)} = 1$ have contribution to $\widetilde{\Gamma}_{B,IF}^{(1)}$, and the sum over the spins $\sigma_i$ and $\sigma_f$ can be calculated independently on the detailed values of $I_{B,\chi}^{(1)}$ and $I_{B,d}^{(1)}$.     
Let us introduce an imaginary-time spin Green's function $D(\mathbf{q},\tau) = -\sum_{ij}\langle T_\tau	S_i (\mathbf{q},\tau) S^{\dag}_j (\mathbf{q},0) \rangle (\delta_{ij} - \widehat{\mathbf{q}}_i \widehat{\mathbf{q}}_j )$. The corresponding spectral function $\chi_B (\mathbf{q},E)$ is defined by $\chi_B (\mathbf{q},E) = -2\, \text{Im}\, D(\mathbf{q},i\nu_n \rightarrow E + i\delta^+)$, which can be shown to follow
\begin{equation}
\chi_B (\mathbf{q},E) = \frac{2\pi}{Z}\sum_{\alpha\beta i j} e^{-\beta E_\alpha} \langle \Psi_\alpha \vert S^\dag_i (\mathbf{q}) \vert \Psi_\beta\rangle \langle \Psi_\beta \vert S_j (\mathbf{q}) \vert \Psi_\alpha\rangle (\delta_{ij} - \widehat{\mathbf{q}}_i \widehat{\mathbf{q}}_j ) n^{-1}_B(E) \delta(E + E_\beta - E_\alpha) , \label{eqnA4.19}
\end{equation}
where $n_B(E)$ is the Bose distribution function. 
Note that $\sum_i S^{(i) \dag}_{\perp}(\mathbf{q})   \, S^{(i)}_{\perp}(\mathbf{q}) = \sum_{i j} S^{\dag}_i (\mathbf{q}) S_j(\mathbf{q}) (\delta_{ij}- \widehat{\mathbf{q}}_i \widehat{\mathbf{q}}_j )$, the statistical average of the single neutron-scattering probability for the INS measurement can be shown to follow 
\begin{equation}
\widebar{\Gamma}_B^{(1)} = \sum_{I F} P_{B}^{(1)}(\mathbf{q}_i) \cdot \Gamma_B^{(1)} \cdot I_{B,\chi}^{(1)} \cdot I_{B,d}^{(1)} ,  \label{eqnA4.20}
\end{equation}
where $\sum_{I F} = \sum_{\mathbf{q}_i n_i n_{f}}$, and $\Gamma_B^{(1)}$ follows   
\begin{equation}
\Gamma_B^{(1)} = \frac{|g(\mathbf{q})|^{2} \Delta t_d}{\hbar} \chi_B(\mathbf{q}, E_B^{(1)}) \cdot n_B(E_B^{(1)}) .  \label{eqnA4.21}
\end{equation}
The transferred energy $E_B^{(1)}$ is defined as above for Eq. (\ref{eqnA4.13}), $E_B^{(1)}=\mathcal{E}(\mathbf{q}_{f}) - \mathcal{E}(\mathbf{q}_{i})$ with $\mathcal{E}(\mathbf{q}_{i})$ and $\mathcal{E}(\mathbf{q}_{f})$ being the incident and the scattered neutron energies. In the simple case with $I_{B,\chi}^{(1)}=1$ and $I_{B,d}^{(1)} = 1$, we can recover the previous result for the INS \citep{SucINS2021} that $\widebar{\Gamma}_B^{(1)} = \Gamma_B^{(1)}$.

\section{$S$-matrix perturbation theory for post-experiment $\text{c}$INS} \label{SecA5}

Let us consider the {\it post-experiment} coincident inelastic neutron scattering (cINS) we have proposed in the main text, which can detect directly the two-spin correlations of the target electrons by coincidence detection of two neutron-scattering processes. The combined system of the {\it post-experiment} cINS measurement is same to that of the INS measurement. Suppose the incident two neutrons from every one of the sequential neutron pulses have momentum and spin distribution functions $P^{(2)}_{B}(\mathbf{q}_{i_1}, \mathbf{q}_{i_2}) = P^{(1)}_{B}(\mathbf{q}_{i_1}) \cdot P^{(1)}_{B}(\mathbf{q}_{i_2})$ and $\widebar{P}^{(2)}_{B}(\sigma_{i_1}, \sigma_{i_2}) = \widebar{P}^{(1)}_{B}(\sigma_{i_1}) \cdot \widebar{P}^{(1)}_{B}(\sigma_{i_2})$. Here $\widebar{P}_{B}^{(1)}(\sigma_{i})$ is defined as in the above INS case with the same neutron-spin mixed states. Suppose the two scattered neutrons arrived at two respective single-neutron detectors $D_1$ and $D_2$ have fixed momenta $(\mathbf{q}_{f_1}, \mathbf{q}_{f_2} )$ but arbitrary spins $(\sigma_{f_1}, \sigma_{f_2})$. The density matrix of the initial states of the {\it pulse}-resolved two neutron-scattering processes of the {\it post}-experiment cINS measurement is defined by 
\begin{equation}
\widetilde{\rho}_B = \sum_{I} P_{B,I}^{(2)} | \Phi_{B,I}^{(2)} \rangle \langle \Phi_{B,I}^{(2)} | , \label{eqnA5.1}
\end{equation}
where $P_{B,I}^{(2)}$ and $| \Phi_{B,I}^{(2)} \rangle$ are defined by
\begin{eqnarray}
&& P_{B,I}^{(2)} = \frac{1}{Z} e^{-\beta E_\alpha} P^{(2)}_{B}(\mathbf{q}_{i_1}, \mathbf{q}_{i_2}) \, \widebar{P}^{(2)}_{B}(\sigma_{i_1}, \sigma_{i_2})  , \label{eqnA5.2-1} \\
&& |\Phi^{(2)}_{B,I} \rangle = | \Psi_{\alpha} \rangle \otimes | n_{\mathbf{q}_{i_1}\sigma_{i_1}} n_{\mathbf{q}_{i_2}\sigma_{i_2}} \rangle . \label{eqnA5.2-2}  
\end{eqnarray}
Here $\sum_{I} = \sum_{\alpha \mathbf{q}_i \sigma_i n_i}$, $n_{\mathbf{q}_{i_1}\sigma_{i_1}}, n_{\mathbf{q}_{i_2}\sigma_{i_2}} = 0, 1$ are two incident neutron numbers of the two respective neutron-scattering processes of each {\it post-experiment} cINS coincidence detection. The projection operator for the final states of the two neutron-scattering processes is defined by
\begin{equation}
1_B^{(2)} = \sum_F |\Phi^{(2)}_{B,F} \rangle \langle \Phi^{(2)}_{B,F} | , \,\, |\Phi^{(2)}_{B,F} \rangle = | \Psi_{\beta} \rangle \otimes | n_{\mathbf{q}_{f_1}\sigma_{f_1}} n_{\mathbf{q}_{f_2}\sigma_{f_2}} \rangle , \label{eqnA5.3}
\end{equation}
where $\sum_F = \sum_{\beta \sigma_f n_f}$, and $n_{\mathbf{q}_{f_1}\sigma_{f_1}}, n_{\mathbf{q}_{f_2}\sigma_{f_2}}=0$ or $1$ are defined for the scattered neutrons which arrive at two single-neutron detectors $D_1$ and $D_2$, respectively. When we introduce the density matrix of the combined system at the observation time $t_f$ as
\begin{equation}
\widetilde{\rho}_{B,I}(t_f) = S_B(t_f, t_i) \widetilde{\rho}_B S_B (t_i, t_f) , \label{eqnA5.4}
\end{equation}   
the coincidence probability of two neutron-scattering processes of the {\it post-experiment} cINS measurement can be defined by
\begin{equation}
\widebar{\Gamma}_B^{(2)} = \text{Tr} [\widetilde{\rho}_{B,I}^{(2)}(t_f) 1_B^{(2)}] = \text{Tr} [ \widetilde{\rho}_B S_B^{(2)}(t_i,t_f) 1_B^{(2)} S_B^{(2)}(t_f, t_i)] , \label{eqnA5.5}
\end{equation}     
where $\widetilde{\rho}_{B,I}^{(2)}(t_f)$ is the second-order part of the density matrix $\widetilde{\rho}_{B,I}(t_f)$, $S_B^{(2)}(t_f,t_i)$ and $S_B^{(2)} (t_i,t_f)$ are the second-order perturbation expansions of the $S_B$-matrices and defined by 
\begin{eqnarray}
&& S_B^{(2)}(t_f, t_i) = \frac{1}{2}(-\frac{i}{\hbar})^2 \iint_{t_i}^{t_f} d t_2 d t_1 T_t [V_{B,I}(t_2) V_{B,I}(t_1)] , \label{eqnA5.6-1} \\
&& S_B^{(2)}(t_f, t_i) = \frac{1}{2}(-\frac{i}{\hbar})^2 \iint_{t_f}^{t_i} d t_1 d t_2 \widetilde{T}_t [V_{B,I}^\dag (t_1) V_{B,I}^\dag (t_2)] . \label{eqnA5.6-2}
\end{eqnarray}
Similarly, $\widebar{\Gamma}_B^{(2)}$ can expressed into the following form as 
\begin{equation}
\widebar{\Gamma}_B^{(2)} = \sum_{I} P_{B,I}^{(2)} \langle \Phi^{(2)}_{B,I} (t_f) | 1_B^{(2)} | \Phi^{(2)}_{B,I}(t_f) \rangle , \, \, |\Phi^{(2)}_{B,I} (t_f)\rangle = S_B^{(2)} (t_f, t_i) | \Phi^{(2)}_{B,I} \rangle . \label{eqnA5.7}
\end{equation}
Thus, the coincidence probability of the {\it post-experiment} cINS measurement can be shown to follow
\begin{equation}
\widebar{\Gamma}_B^{(2)}=\frac{1}{Z}\sum_{I F} e^{-\beta E_\alpha} P^{(2)}_{B}(\mathbf{q}_{i_1}, \mathbf{q}_{i_2}) \, \widebar{P}^{(2)}_{B}(\sigma_{i_1}, \sigma_{i_2}) \big| \langle \Phi^{(2)}_{B,F} | S_B^{(2)} (t_f, t_i) | \Phi^{(2)}_{B,I} \rangle \big|^2 , \label{eqnA5.8} 
\end{equation}
where $\sum_{IF}=\sum_{\alpha\beta \mathbf{q}_i \sigma_i \sigma_f n_i n_f}$. Eq. (\ref{eqnA5.8}) is one main result of the $S$-matrix perturbation theory for the {\it post-experiment} cINS.  

Let us consider the coincidence detection of two neutron-scattering processes from one neutron pulse with one initial state $|\Phi^{(2)}_{B,I} \rangle$ and one final state $|\Phi^{(2)}_{B,F} \rangle$. The coincidence probability of the two neutron-scattering processes is  defined by 
\begin{equation}
\widebar{\Gamma}_{B,IF}^{(2)} = \big| \langle \Phi^{(2)}_{B,F} | S_B^{(2)} (t_f, t_i)  | \Phi^{(2)}_{B,I} \rangle \big|^2 . \label{eqnA5.9}
\end{equation}
$\widebar{\Gamma}_{B,IF}^{(2)}$ can be calculated by  
\begin{eqnarray}
\hspace{-2em} \widebar{\Gamma}_{B,IF}^{(2)} &=& \Big| \frac{1}{2} (-\frac{i}{\hbar})^2 \iint_{t_i}^{t_f} d t_2 d t_1  \langle n_{\mathbf{q}_{f_1}\sigma_{f_1}} n_{\mathbf{q}_{f_2}\sigma_{f_2}}; \Psi_\beta | T_t [V_{B,I}(t_2) V_{B,I}(t_1)] | \Psi_\alpha; n_{\mathbf{q}_{i_1}\sigma_{i_1}} n_{\mathbf{q}_{i_2}\sigma_{i_2}} \rangle  \Big|^2 \notag \\
&=& \Big| \frac{1}{2} (-\frac{i}{\hbar})^2 \iint_{t_i}^{t_f} d t_2 d t_1  \sum_{\mathbf{q}_i^\prime \mathbf{q}_f^\prime \sigma_i^\prime \sigma_f^\prime} \sum_{ij} g_B(\mathbf{q}_1^\prime) g_B(\mathbf{q}_2^\prime) \langle \Psi_\beta | S^{(j)}_{\perp}(\mathbf{q}_2^\prime, t_2) S^{(i)}_{\perp}(\mathbf{q}_1^\prime, t_1)  | \Psi_\alpha \rangle \, \boldsymbol{\tau}^{j}_{\sigma_{f_2}^\prime \sigma_{i_2}^\prime} \boldsymbol{\tau}^{i}_{\sigma_{f_1}^\prime \sigma_{i_1}^\prime} \, \Big. \notag \\
&& \times \Big. \langle n_{\mathbf{q}_{f_1}\sigma_{f_1}} n_{\mathbf{q}_{f_2}\sigma_{f_2}} | f^{\dag}_{\mathbf{q}_{f_2}^\prime \sigma_{f_2}^\prime} (t_2) f^{\dag}_{\mathbf{q}_{f_1}^\prime \sigma_{f_1}^\prime} (t_1) | 0 \rangle  \langle 0 | f_{\mathbf{q}_{i_2}^\prime \sigma_{i_2}^\prime} (t_2) f_{\mathbf{q}_{i_1}^\prime \sigma_{i_1}^\prime} (t_1) | n_{\mathbf{q}_{i_1}\sigma_{i_1}} n_{\mathbf{q}_{i_2}\sigma_{i_2}} \rangle  \Big|^2 \notag \\
&=& \widebar{\Gamma}_{B,IF}^{(2,1)} + \widebar{\Gamma}_{B,IF}^{(2,2)} . \label{eqnA5.10}
\end{eqnarray}
$\widebar{\Gamma}_{B,IF}^{(2,1)}$ and $\widebar{\Gamma}_{B,IF}^{(2,2)}$ define the two contributions from two different classes of microscopic neutron-scattering processes. $\widebar{\Gamma}_{B,IF}^{(2,1)}$ comes from the neutron-scattering processes with the neutron-state changes as $\vert \mathbf{q}_{i_1}\sigma_{i_1}\rangle \rightarrow \vert \mathbf{q}_{f_1}\sigma_{f_1}\rangle$ and $\vert \mathbf{q}_{i_2}\sigma_{i_2}\rangle \rightarrow \vert \mathbf{q}_{f_2}\sigma_{f_2}\rangle$, and $\widebar{\Gamma}_{B,IF}^{(2,2)}$ stems from the neutron-state changes as $\vert \mathbf{q}_{i_1}\sigma_{i_1}\rangle \rightarrow \vert \mathbf{q}_{f_2}\sigma_{f_2}\rangle$ and $ \vert \mathbf{q}_{i_2}\sigma_{i_2}\rangle \rightarrow \vert \mathbf{q}_{f_1}\sigma_{f_1}\rangle$. Let us introduce a two-spin Bethe-Salpeter wave function for the target-electron spin system \citep{SucINS2021}, 
\begin{equation}
\phi^{(ij)}_{\alpha\beta}(\mathbf{q}_1 t_1, \mathbf{q}_2 t_2) = \langle \Psi_\beta \vert T_t S^{(j)}_\perp (\mathbf{q}_2, t_2) S^{(i)}_\perp (\mathbf{q}_1, t_1) \vert \Psi_\alpha \rangle . \label{eqnA5.11}
\end{equation}
With a similar treatment for the cARPES, we can define a center-of-mass time $t_c = (t_1 + t_2)/2$ and a relative time $t_r = t_2 - t_1$. The two-spin Bethe-Salpeter wave function can be reexpressed into the form as $\phi^{(ij)}_{\alpha\beta}(\mathbf{q}_1, \mathbf{q}_2; t_c, t_r ) = \phi^{(ij)}_{\alpha\beta}(\mathbf{q}_1 t_1, \mathbf{q}_2 t_2)$. The frequency Fourier transformation form $\phi^{(ij)}_{\alpha\beta}(\mathbf{q}_1 , \mathbf{q}_2; \Omega,\omega)$ can be defined by
\begin{equation}
\phi^{(ij)}_{\alpha\beta}(\mathbf{q}_1, \mathbf{q}_2; \Omega, \omega) = \iint^{+\infty}_{-\infty} d t_c d t_r \phi^{(ij)}_{\alpha\beta}(\mathbf{q}_1, \mathbf{q}_2; t_c, t_r) e^{i\Omega t_c + i\omega t_r} . \label{eqnA5.12} 
\end{equation}
In the limit with $t_i = \rightarrow -\infty$ and $t_f = \rightarrow +\infty$, it can be shown that   
\begin{eqnarray}
\widebar{\Gamma}_{B,IF}^{(2,1)}
&=& \widebar{\Gamma}_{B,\alpha\beta}^{(2,1)}  \cdot I_{B,\chi}^{(2)} \cdot I_{B,d}^{(2)} , \label{eqnA5.13-1} \\
\widebar{\Gamma}_{B,IF}^{(2,2)} &=& \widebar{\Gamma}_{B,\alpha\beta}^{(2,2)} \cdot I_{B,\chi}^{(2)} \cdot I_{B,d}^{(2)} , \label{eqnA5.13-2}
\end{eqnarray}
where $\widebar{\Gamma}_{B,\alpha\beta}^{(2,1)}$ and $\widebar{\Gamma}_{B,\alpha\beta}^{(2,2)}$ are given by
\begin{eqnarray}
\widebar{\Gamma}_{B,\alpha\beta}^{(2,1)} &=& \frac{|g_B(\mathbf{q}_1) g_B(\mathbf{q}_2)|^2}{\hbar^4}  \big| \sum_{ij} \phi^{(ij)}_{\alpha\beta}(\mathbf{q}_1 , \mathbf{q}_2; \Omega_B,\omega_B) \boldsymbol{\tau}^{j}_{\sigma_{f_2} \sigma_{i_2}} \boldsymbol{\tau}^{i}_{\sigma_{f_1} \sigma_{i_1}} \big|^2 ,  \label{eqnA5.14} \\
\widebar{\Gamma}_{B,\alpha\beta}^{(2,2)} &=& \frac{|g_B(\overline{\mathbf{q}}_1) g_B(\overline{\mathbf{q}}_2)|^2}{\hbar^4}  \big| \sum_{ij} \phi^{(ij)}_{\alpha\beta}(\overline{\mathbf{q}}_1 , \overline{\mathbf{q}}_2; \overline{\Omega}_B, \overline{\omega}_B) \boldsymbol{\tau}^{j}_{\sigma_{f_2} \sigma_{i_1}} \boldsymbol{\tau}^{i}_{\sigma_{f_1} \sigma_{i_2}} \big|^2  . \label{eqnA5.15}
\end{eqnarray} 
$I_{B,\chi}^{(2)}$ is an incident-neutron-state factor and $I_{B,d}^{(2)}$ is a scattered-neutron-state factor,  which are defined for the {\it post-experiment} cINS detection as
\begin{equation}
I_{B,\chi}^{(2)} = \big| \langle 0 | f_{\mathbf{q}_{i_1}\sigma_{i_1}} | n_{\mathbf{q}_{i_1}\sigma_{i_1}} \rangle \big|^2 \cdot \big| \langle 0 | f_{\mathbf{q}_{i_2}\sigma_{i_2}} | n_{\mathbf{q}_{i_2}\sigma_{i_2}} \rangle \big|^2 , \label{eqnA5.16}
\end{equation}
and 
\begin{equation}
I_{B,d}^{(2)} = I_{B,d_1}^{(1)} \times I_{B,d_2}^{(1)} , \label{eqnA5.17}
\end{equation} 
where $I_{B,d_1}^{(1)}$ and $I_{B,d_2}^{(1)}$ are defined by
\begin{equation}
I_{B,d_1}^{(1)} = \big| \langle n_{\mathbf{q}_{f_1}\sigma_{f_1}} | f^\dag_{\mathbf{q}_{f_1}\sigma_{f_1}} | 0 \rangle \big|^2, \,\, I_{B,d_2}^{(1)} = \big| \langle n_{\mathbf{q}_{f_2}\sigma_{f_2}} | f^\dag_{\mathbf{q}_{f_2}\sigma_{f_2}} | 0 \rangle \big|^2 . \label{eqnA5.18}
\end{equation} 
It should be noted that $I_{B,d_1}^{(1)} = 0 \, (1)$ when $n_{\mathbf{q}_{f_1}\sigma_{f_1}} = 0 \, (1)$ and $I_{B,d_2}^{(1)} = 0 \, (1)$ when $n_{\mathbf{q}_{f_2}\sigma_{f_2}} = 0 \, (1)$. Therefore, $I_{B,d}^{(2)}$ records the coincidence counting of the scattered neutrons arrived at two single-neutron detectors.  
In Eqs. (\ref{eqnA5.14}) and (\ref{eqnA5.15}),  the transferred momenta are defined by 
\begin{equation}
\mathbf{q}_1 = \mathbf{q}_{f_1}-\mathbf{q}_{i_1} , \mathbf{q}_2 = \mathbf{q}_{f_2}-\mathbf{q}_{i_2}, \overline{\mathbf{q}}_1 = \mathbf{q}_{f_1}-\mathbf{q}_{i_2} , \overline{\mathbf{q}}_2 = \mathbf{q}_{f_2}-\mathbf{q}_{i_1} , \label{eqnA5.19} 
\end{equation}
and the transferred frequencies are defined by 
\begin{equation}
\Omega_B = \frac{1}{\hbar}(E_{B,1} + E_{B,2} ), \omega_B = \frac{1}{2\hbar}(E_{B,2} - E_{B,1} ), \overline{\Omega}_B = \frac{1}{\hbar}(\overline{E}_{B,1} + \overline{E}_{B,2} ), \overline{\omega}_B = \frac{1}{2\hbar}(\overline{E}_{B,2} - \overline{E}_{B,1} ) . \label{eqnA5.20}
\end{equation}
Here the transferred energies in two relevant neutron-scattering processes of the {\it post-experiment} cINS detection are defined as 
\begin{equation}
E_{B,1} = \mathcal{E}(\mathbf{q}_{f_1}) - \mathcal{E}(\mathbf{q}_{i_1}) , E_{B,2} = \mathcal{E}(\mathbf{q}_{f_2}) - \mathcal{E}(\mathbf{q}_{i_2}) , \overline{E}_{B,1} = \mathcal{E}(\mathbf{q}_{f_1}) - \mathcal{E}(\mathbf{q}_{i_2}) ,  \overline{E}_{B,2} = \mathcal{E}(\mathbf{q}_{f_2}) - \mathcal{E}(\mathbf{q}_{i_1}) . \label{eqnA5.21}
\end{equation}
In summary, $\widebar{\Gamma}_{B,IF}^{(2)}$ can be expressed into the form as  
\begin{equation}
\widebar{\Gamma}_{B,IF}^{(2)} = \widebar{\Gamma}_{B,\alpha\beta}^{(2)} \cdot I_{B,\chi}^{(2)} \cdot I_{B,d}^{(2)} ,  \label{eqnA5.22}
\end{equation}
where $\widebar{\Gamma}_{B,\alpha\beta}^{(2)}$ is defined as
\begin{equation}
\widebar{\Gamma}_{B,\alpha\beta}^{(2)} = \widebar{\Gamma}_{B,\alpha\beta}^{(2,1)} + \widebar{\Gamma}_{B,\alpha\beta}^{(2,2)} . \label{eqnA5.23}
\end{equation}
Here $\widebar{\Gamma}_{B,\alpha\beta}^{(2,1)}$ and $\widebar{\Gamma}_{B,\alpha\beta}^{(2,2)}$ are given by Eqs. (\ref{eqnA5.14}) and (\ref{eqnA5.15}). It should be remarked that in the derivation of the two contributions from two different classes of two neutron-scattering processes, we have ignored the quantum interference from these two different classes of neutron-scattering processes. 

The statistical average of the coincidence probability of {\it pulse-resolved} two neutron-scattering processes from every one of the sequential neutron pulses can be calculated from Eq. (\ref{eqnA5.8}), which follows
\begin{equation}
\widebar{\Gamma}_B^{(2)}=\frac{1}{Z}\sum_{I F} e^{-\beta E_\alpha} P^{(2)}_{B}(\mathbf{q}_{i_1}, \mathbf{q}_{i_2}) \widebar{P}^{(2)}_{B}(\sigma_{i_1}, \sigma_{i_2}) \cdot \widebar{\Gamma}_{B,\alpha\beta}^{(2)} \cdot I_{B,\chi}^{(2)} \cdot I_{B,d}^{(2)} , \label{eqnA5.24} 
\end{equation}
where $\sum_{IF}=\sum_{\alpha\beta \mathbf{q}_i \sigma_i \sigma_f n_i n_f}$. With a same method for the sum over the spins $\sigma_{i}$ and $\sigma_{f}$ as used in Eq. (\ref{eqnA4.18}), $\widebar{\Gamma}_B^{(2)}$ can be shown to follow
\begin{equation}
\widebar{\Gamma}_B^{(2)} = \sum_{I F} P^{(2)}_{B}(\mathbf{q}_{i_1},\mathbf{q}_{i_2}) \cdot \Gamma_{B}^{(2)} \cdot I_{B,\chi}^{(2)} \cdot I_{B,d}^{(2)} ,  \label{eqnA5.25}
\end{equation}
where $\sum_{I F} = \sum_{\mathbf{q}_{i_1} \mathbf{q}_{i_2}} \sum_{ n_{i_1} n_{i_2} n_{f_1} n_{f_2} }$, and $\Gamma_{B}^{(2)}$ follows \cite{SucINS2021} 
\begin{equation}
\Gamma_{B}^{(2)} = \Gamma_{B,1}^{(2)} + \Gamma_{B,2}^{(2)} , \label{eqnA5.26}
\end{equation}
with the two contributions defined as
\begin{eqnarray}
\Gamma_{B,1}^{(2)} &=& \frac{1}{Z}\sum_{\alpha\beta i j} e^{-\beta E_{\alpha}} C_1 \big\vert \phi^{(ij)}_{\alpha\beta}(\mathbf{q}_1 , \mathbf{q}_2; \Omega_B,\omega_B)  \big\vert^2 , \label{eqnA5.27} \\
\Gamma_{B,2}^{(2)} &=& \frac{1}{Z}\sum_{\alpha\beta i j} e^{-\beta E_{\alpha}} C_2 \big\vert \phi^{(ij)}_{\alpha\beta}(\overline{\mathbf{q}}_1 , \overline{\mathbf{q}}_2; \overline{\Omega}_B, \overline{\omega}_B) \big\vert^2 .  \label{eqnA5.28} 
\end{eqnarray}
Here the two constants $C_1$ and $C_2$ are given by $C_1 = \vert g_B(\mathbf{q}_1) g_B(\mathbf{q}_2) \vert^2 /\hbar^4$ and $C_2 = \vert g_B(\overline{\mathbf{q}}_1) g_B(\overline{\mathbf{q}}_2) \vert^2/\hbar^4$. It should be noted that $I_{B,d}^{(2)}= I_{B,d_1}^{(1)} \times I_{B,d_2}^{(1)}$ follows in $\widebar{\Gamma}_{B,IF}^{(2)}$ and $\widebar{\Gamma}_{B}^{(2)}$. Therefore, the coincidence probability of the {\it post-experiment} cINS measurement can be obtained by $I_{B,d}^{(2)}$ which records the coincidence counting of the scattered neutrons from {\it pulse}-resolved two neutron-scattering processes with the renormalization of the target-electron spin form factor and the incident-neutron-state factor. When we consider the case with $I_{B,\chi}^{(2)}=1$ and $I_{B,d}^{(2)}=1$, we can recover our previous results for the {\it instantaneous} cINS \cite{SucINS2021}.

From the coincidence probability of the {\it post-experiment} cINS measurement in Eq. (\ref{eqnA5.25}), it is clear that the {\it post-experiment} cINS can provide the information on the frequency two-spin Bethe-Salpeter wave function. Therefore, it will be a powerful technique to study the dynamical two-spin correlations of the target electrons. This can be seen more clearly from the following spectrum expression of the frequency two-spin Bethe-Salpeter wave function \citep{SucINS2021}: 
\begin{equation} 
\phi^{(ij)}_{\alpha\beta}\left(\mathbf{q}_1, \mathbf{q}_2; \Omega, \omega \right)  = 2\pi \delta \left[\Omega + \left( E_{\beta} - E_{\alpha}\right)/\hbar \right] \phi^{(ij)}_{\alpha\beta}\left(\mathbf{q}_1, \mathbf{q}_2; \omega \right) , \label{eqnA5.29} 
\end{equation} 
where $\phi^{(ij)}_{\alpha\beta}\left(\mathbf{q}_1, \mathbf{q}_2; \omega \right)$ follows
\begin{equation}
\phi^{(ij)}_{\alpha\beta}\left(\mathbf{q}_1, \mathbf{q}_2; \omega \right) =\sum_{\gamma} \left[ \frac{ i \langle\Psi_{\beta} \vert S^{(j)}_\perp (\mathbf{q}_2)  \vert \Psi_{\gamma} \rangle \langle \Psi_{\gamma} \vert  S^{(i)}_\perp (\mathbf{q}_1) \vert \Psi_{\alpha} \rangle} {\omega + i\delta^+ + (E_{\alpha} + E_{\beta} - 2 E_{\gamma} )/2\hbar} - \frac{ i \langle\Psi_{\beta} \vert S^{(i)}_\perp (\mathbf{q}_1) \vert \Psi_{\gamma} \rangle \langle \Psi_{\gamma} \vert  S^{(j)}_\perp (\mathbf{q}_2) \vert \Psi_{\alpha} \rangle} {\omega - i\delta^+ - (E_{\alpha} + E_{\beta} - 2 E_{\gamma} )/2\hbar} \right] . \label{eqnA5.30}
\end{equation}
Obviously, the {\it post-experiment} cINS can provide the dynamical two-spin correlations of the target electrons, which involve both the center-of-mass and the inner-pair relative dynamical physics with both momentum and energy resolved.



%